\documentclass[journal]{IEEEtran}
\usepackage{cite}
\usepackage{amsmath,amssymb,amsfonts}
\usepackage{algorithm}
\usepackage[noend]{algpseudocode}
\usepackage{graphicx}
\usepackage{textcomp}

\usepackage{subcaption}

\usepackage{amssymb}
\usepackage{multirow}
\usepackage{float}
\usepackage{hyperref}
\usepackage{xcolor}

\ifCLASSINFOpdf
\else
\fi
\begin{document}
%\title{Self-Supervised Contrastive Learning-based GNN for real-time Seismic Intensity Map Generation}
\title{Real-time Seismic Intensity Prediction using Self-supervised Contrastive GNN for Earthquake Early Warning}
%Rafid Umayer Murshed, Kazi Noshin, Md. Anu Zakaria, Md. Forkan Uddin, A. F. M. S. Amin, Mohammed Eunus Ali
\author{Rafid~Umayer~Murshed,~\IEEEmembership{Member,~IEEE,}
        Kazi~Noshin,
        Md.~Anu~Zakaria,\\
        Md.~Forkan~Uddin,~\IEEEmembership{Member,~IEEE,}
        A. F. M. Saiful Amin,
        and~Mohammed~Eunus~Ali$^{\ast}$
\thanks{R. U. Murshed (e-mail: rafid.buet.eee16@gmail.com), K. Noshin, and M. A. Zakaria are with the BUET-Japan Institute of Disaster Prevention and Urban Safety (JIDPUS), Bangladesh University of Engineering and Technology, Dhaka 1000.}% <-this % stops a space
\thanks{M. F. Uddin is with the Department of Electrical and Electronics Engineering, Bangladesh University of Engineering and Technology, Dhaka 1000.}% <-this % stops a space
\thanks{A. F. M. S. Amin is with the Department of Civil Engineering and BUET-JIDPUS, Bangladesh University of Engineering and Technology, Dhaka 1000.}% <-this % stops a space
\thanks{M. E. Ali is with the Department of Computer Science and Engineering, Bangladesh University of Engineering and Technology, Dhaka 1000. e-mail: eunus@cse.buet.ac.bd.}% <-this % stops a space
\thanks{$^{\ast}$The corresponding author.}
}


\markboth{IEEE Transactions on Geoscience and Remote Sensing,~Vol.~xx, No.~x, February~2024}%
{Shell \MakeLowercase{\textit{Murshed et al.}}: Bare Demo of IEEEtran.cls for Journals}



\maketitle

\begin{abstract}
%\Eunus{modified the abstract, rafid, pls check}
Seismic intensity prediction from early or initial seismic waves received by a few seismic stations can enhance Earthquake Early Warning (EEW) systems, particularly in ground motion-based approaches like PLUM. While many operational EEW systems currently utilize point-source-based models that estimate the warning area based on magnitude and distance measures, direct intensity prediction offers a potential improvement in accuracy and reliability. In this paper, we propose a novel deep learning approach, Seismic Contrastive Graph Neural Network (SC-GNN), for highly accurate seismic intensity prediction using a small portion of initial seismic waveforms from a few seismic stations. The SC-GNN consists of two key components: (i) a graph neural network (GNN) to propagate spatiotemporal information through a graph-like structure representing seismic station distribution and wave propagation, and (ii) a self-supervised contrastive learning component to train the network with larger time windows and enable predictions using shorter initial waveforms. The efficacy of our approach is demonstrated through experiments on three real-world seismic datasets, where it shows superior performance over existing techniques, including a significant reduction in mean squared error (MSE) and the lowest standard deviation of error, indicating its robustness, reliability, and strong positive relationship between predicted and actual values. Notably, the SC-GNN model maintains superior performance even with 5s input waveforms, making it especially suitable for enhancing EEW applications.

\end{abstract}


% \fi
\IEEEpeerreviewmaketitle

\section{Introduction}

\IEEEPARstart{E}{arthquakes} represent a major threat to lives and property, as exemplified by the devastating M7.8 Turkey-Syria earthquake on February 6, 2023. This event caused over fifty thousand fatalities and inflicted extensive economic, infrastructural, and cultural heritage damage \cite{ni2023complexities}. While Earthquake Early Warning (EEW) systems cannot prevent earthquake damages, they are crucial in reducing fatalities by providing advance warning \cite{kanamori2003}. EEW systems, which detect initial seismic waves (P-waves), offer critical seconds to minutes of warning before the more destructive waves (S and surface waves) arrive. However, it is important to note that most operational EEW systems use point-source-based models, estimating the warning area based on magnitude and distance, rather than direct intensity prediction. Ground motion-based EEW systems like PLUM, which rely on seismic intensity prediction, are essential for forecasting an earthquake's potential impact at different geographical locations \cite{wald1999relationships}. This critical predictive element plays a pivotal role in delivering actionable information, facilitating prompt safety measures to alleviate the repercussions of seismic events on human life and infrastructure.

The history of EEW systems and seismic intensity prediction spans several decades, with initial efforts focused on detecting early seismic waves like P-waves to provide short lead times for warnings. Despite advancements in seismic monitoring networks and computational techniques, it is important to note that almost all operational EEW systems worldwide currently rely on point-source-based models. More sophisticated methods for seismic intensity prediction, including physics-based models, empirical relationships, and machine learning algorithms, are predominantly in the research phase (e.g.,\cite{allen2019earthquake,mousavi2020machine}).

Traditional methods for seismic prediction tasks have primarily utilized empirical models and classical machine learning techniques. Ground motion prediction equations (GMPEs), which are often empirical models derived from fitting data to strong motion records, are widely used to estimate ground motion parameters like PGA and SA, considering source, path, and site parameters \cite{boore1997equations}. Classical machine learning techniques, such as support vector machines (SVMs) \cite{astuti2014hybrid} and artificial neural networks (ANNs) \cite{erdik2009artificial,chen2022automated}, have been explored to estimate ground motion parameters or predict seismic events. For a comprehensive understanding of different machine learning approaches for GMPE building, readers are referred to \cite{mousavi2023machine}.

\begin{figure*}[! h]
\centering
    \includegraphics[width=2\columnwidth,height=85mm]{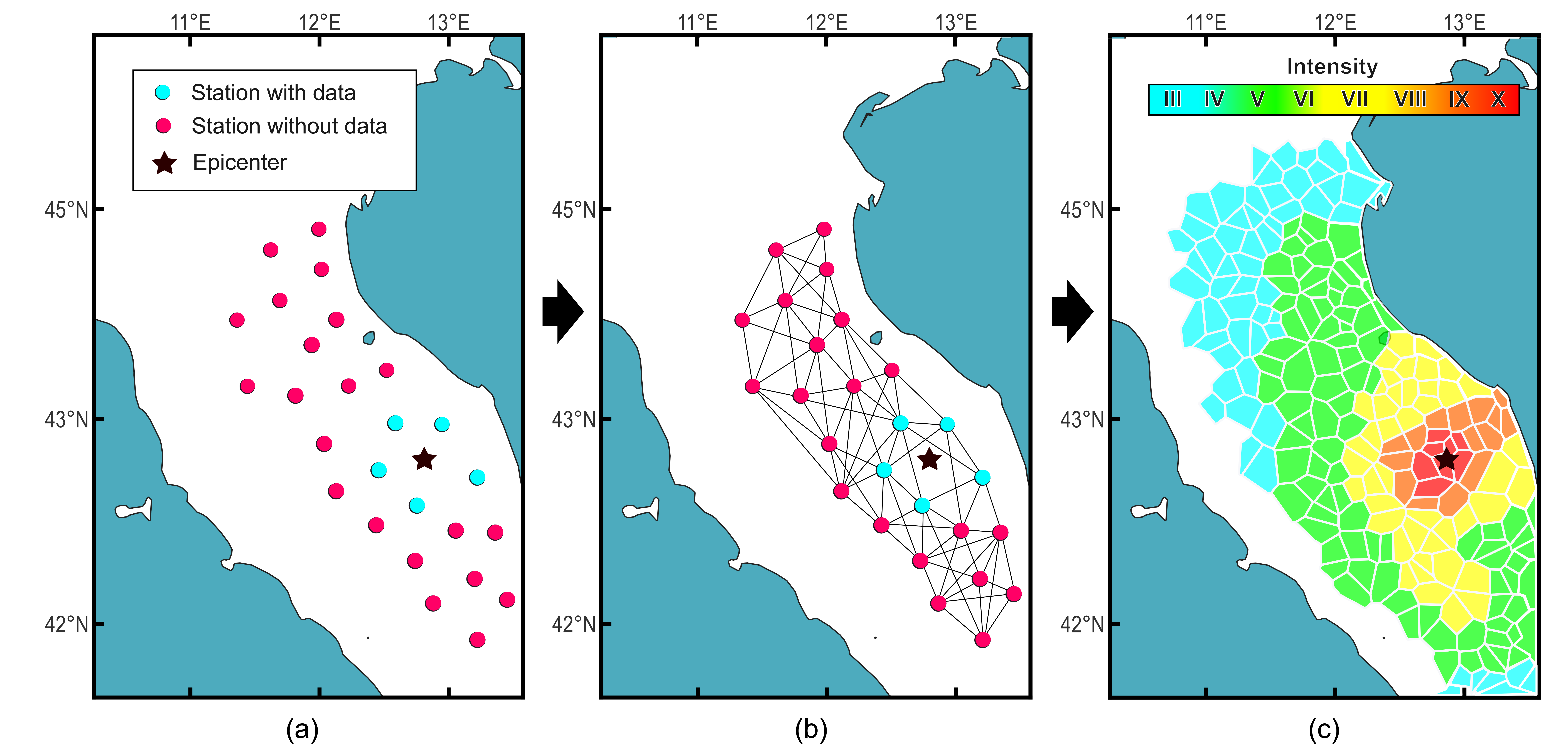}
    \caption{Seismic intensity prediction with SC-GNN}
    \vspace{-2mm}
    \label{Intro_Fig}
\end{figure*}

Deep learning has gained popularity in various domains in recent years due to its ability to learn complex representations from large datasets \cite{DONG2021100379,self_HBF, self_ALC}. Recent advancements in deep learning have significantly influenced seismic data processing and model building in seismology \cite{mousavi2022deep}. Its ability to learn complex representations from large datasets makes it particularly suitable for processing seismic data, offering new avenues for understanding and predicting seismic phenomena. For instance, deep learning has been applied to earthquake detection \cite{perol2018convolutional}, magnitude estimation \cite{mousavi2020machine}, and location and origin time estimation \cite{cnn_loc_origin,cnn_loc_origin_2}. These applications include the use of convolutional neural networks (CNNs), recurrent neural networks (RNNs), and other architectures. Notably, \cite{mousavi2022deep,mousavi2023machine} provide comprehensive overviews of various deep learning approaches in this field, highlighting their potential and limitations. Generative Adversarial Networks (GAN) have been used to reduce false triggers from local noises \cite{GAN}. Moreover, deep learning techniques have been employed in EEW systems and seismic intensity prediction tasks, demonstrating promising results in terms of accuracy and speed. In \cite{ground_motion_intensity}, a CNN-based technique has been used to predict ground motion intensity due to earthquakes using a 10s window from the earthquake origin time without requiring prior knowledge of earthquake sources. Transfer learning has also been used to predict ground shaking of an area due to earthquake \cite{transfer_learning}. A graph convolutional network (GCN) based approach, namely  TISER-GCN, for multivariate time-series regression is proposed in \cite{bloemheuvel2022graph}. This approach is tested on two seismic datasets for intensity prediction and demonstrates promising results with an average MSE reduction of 16.3\%— an improvement compared to \cite{ground_motion_intensity}.

The ROSERS framework proposed in \cite{Rosers} prioritizes \emph{on-site EEW} using DNN and variational autoencoders. Early p-waves and site features are used to estimate the acceleration response spectrum (Sa(T)) of ground-motion waveforms at the station of interest. Instead of using waveforms from local stations to predict intensities at remote sites, ROSERS requires data from the seismic station being studied. In broad geographic early warning scenarios, this localized approach may be inadequate. In \cite{HSU2013210}, Support Vector Regression (SVR) is used to estimate PGA from P-wave features collected from vertical ground acceleration at a station. It evaluates its on-site prediction model using previous earthquake data. This approach can underestimate PGA for earthquakes with complex slip propagation and is also confined to predictions at the data station. The transformer-based TEAM model \cite{10.1093/gji/ggaa609} focuses on EEW by capturing sequential patterns in seismic data; however, it predicts intensity distributions rather than specific values. This approach suffers from high computational complexity and can not adequately handle the spatiotemporal characteristics inherent to seismic data for highly accurate seismic intensity prediction at geographically distant locations, as found in our empirical evaluation.

The effectiveness of seismic intensity prediction algorithms varies significantly between point-source EEW systems and ground-motion-based EEW systems. While point-source EEW systems estimate the earthquake's location, depth, and magnitude to infer intensity, ground-motion-based systems, such as PLUM, directly estimate the upcoming intensity using network-wide ground motion observations, bypassing estimations of magnitude and location. This distinction is crucial for understanding the inherent limitations of each approach. For instance, traditional methods, including physics-based modeling and classical machine learning techniques, may not adequately address the complexities of seismic data, leading to challenges in real-time accuracy for both types of systems\cite{espinosa1995mexico}. Furthermore, while deep learning has advanced seismic intensity prediction, achieving optimal performance remains a challenge due to the need for large time windows and the intricate nature of seismic data \cite{transfer_learning,ground_motion_intensity,bloemheuvel2022graph}. The precision of intensity predictions is critical; even minor inaccuracies can lead to significant consequences, such as over or under-alerting. In EEW systems, the estimated Modified Mercalli Intensity (MMI) level determines whether to issue an alert and the scope of the alerted area. An underestimation by even one unit of MMI could result in a failure to trigger alerts for events that should have been recognized. Furthermore, as the damage caused by an earthquake increases by a factor of 10 for each 1.0-point rise in intensity scale \cite{wald1999relationships}, precise predictions are vital for effective earthquake preparedness and response efforts. Moreover, every second saved in early warning generation can increase the alerted area by a few kilometers, protecting many more lives and properties \cite{franchi2022can}. 

%Early prediction involves leveraging the initial seismic waveforms received by proximal seismic stations, extracting critical earthquake information from these, and using it to forecast the seismic intensities at an array of stations spanning the affected region. 

Early prediction involves leveraging the initial seismic waveforms received by proximal seismic stations, extracting critical earthquake features from these initial waveforms, and using these features to forecast the seismic intensities at an array of stations spanning the affected region. Informed by the limitations of existing EEW systems and seismic intensity prediction algorithms, we present an approach that utilizes a relatively small segment of initial seismic waveforms from multiple seismic stations dispersed across a geographically sparse region. We aim to accurately predict seismic intensity at these stations and others in the surrounding area where the seismic waves have not yet arrived \cite{allen2009real}. Recognizing the graph-like structure of seismic station distribution and their interaction with seismic wavefield propagation, we employ a graph neural network (GNN) as the foundation of our approach \cite{rong2020deep, yao2019graph}. The unique power of GNNs lies in their ability to propagate information through the nodes of a graph, allowing us to predict seismic intensity at distant stations by exploiting a fraction of the information gathered at the early-receiving stations. In essence, GNNs enable us to make globally informed predictions with locally available data. Most importantly, to address the need for shorter time-window predictions, we incorporate self-supervised contrastive learning, enabling the model to be trained using larger time windows while making predictions using shorter ones \cite{chen2020simple,he2020momentum}. This integration of contrastive learning and specialized GNN layers results in an effective and efficient approach that can perform on significantly shorter time windows. Moreover, the inherent self-supervised nature of this contrastive learning approach eliminates the necessity for exhaustive labeling of the input data. We aptly name our proposed model as Seismic Contrastive GNN (SC-GNN).

We have demonstrated the efficacy of our approach through a comprehensive series of experiments conducted using three real-world seismic datasets. Experimental results substantiate that our approach consistently surpasses the performance of state-of-the-art techniques across a broad spectrum of evaluation metrics. In particular, on our principal dataset, our SC-GNN model demonstrates substantial improvement with a mean squared error (MSE) of 0.4172, reflecting an approximately 234\% enhancement over the best-performing state-of-the-art GCN model. Additionally, our model maintains the lowest standard deviation of error of 0.61 and attains the highest correlation coefficient, indicating robustness, reliability, and a strong positive relationship between predicted and actual values. As the input time window diminishes, our model's performance remains consistently superior to the baseline models, underlining its capability to handle variable input sizes efficiently.
The main contributions of this paper are threefold:

\begin{enumerate}

\item  We propose a contrasting learning-based deep learning framework for near real-time seismic intensity prediction that facilitates highly accurate seismic intensity prediction from much shorter input waveforms than its competitive methods. Numerical results demonstrate that our model achieves superior performance (143\% improvement) with an input window length of 5s compared to the 10s window input of the other baseline models. Also, the self-supervised nature of the proposed framework eliminates the need for extensive labeled data during the contrastive training phase. 

\item  We construct a cutting-edge GNN architecture comprising highly sophisticated graph convolutional and attention layers that capture the complex spatial relationships between seismic stations. This allows us to effectively model the propagation of seismic waves received from a geographically sparse set of seismic stations. 

\item We present the SC-GNN model's exceptional performance compared to baseline models across all standard metrics. The superior performance is reflected in a substantially reduced Mean Squared Error (MSE) of 0.41 on the primary dataset, a lower standard deviation of the error indicating higher reliability, and a high correlation coefficient of around 84\%, suggesting a robust match between predicted and actual values. Furthermore, the SC-GNN model promises remarkable utility in EEW systems, where approximately 70\% of the locations potentially receive a warning time of more than 10 seconds, sufficient for taking various precautionary measures.

\end{enumerate}

The remainder of this paper is organized as follows. Section II formulates the problem. Section III provides a brief background on graph neural networks and contrastive learning. Section IV describes the proposed method, including the GNN architecture and the self-supervised contrastive learning framework. Section V presents the experimental setup, results, and comparisons with existing approaches. Finally, Section VI concludes the paper and discusses potential future directions.

\section{Problem Formulation}

\begin{figure*}[! h]
\centering
    \includegraphics[width=1.9\columnwidth,height=55mm]{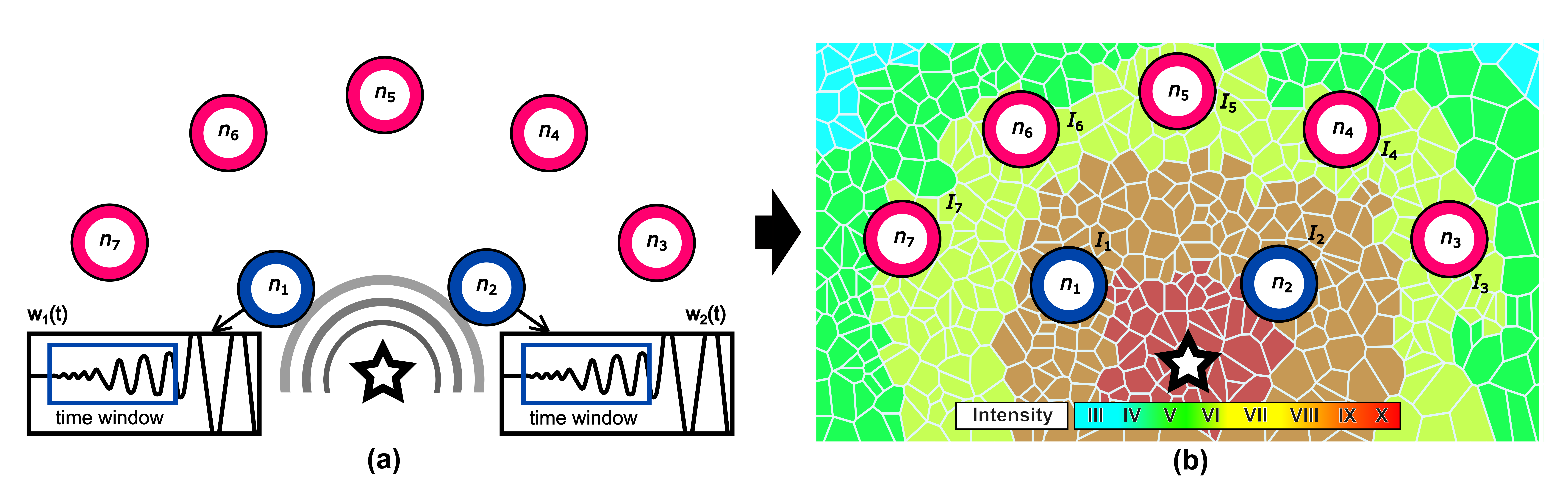}
    \caption{ a) Earthquake occurrence and the propagation of seismic waves b) Seismic intensity prediction at seismic stations and points of interest.}
    \vspace{-2mm}
    \label{Probfor}
\end{figure*}

One of the key challenges of earthquake early warning (EEW) systems is to predict the seismic intensity in surrounding geographical locations at the earliest possible time. As seismic stations are placed in geographically sparse locations in a region of interest, initial seismic waves may only be detected by the few seismic stations located nearby near the earthquake's epicenter. Thus, given the few seismic stations have detected the initial seismic waves, our goal in this paper is to accurately predict the seismic intensity for all the points of interest, including the distant seismic stations within a geographical region of our interest.

Let $\mathcal{N}=\{n_1, n_2, ...n_{|\mathcal{N}|}\}$  be the set of $N$ stations (seismic stations and points of interest). Also, let $\mathcal{N}^{\prime} \subset \mathcal{N}$, $\mathcal{|N}^{\prime}| \lll |\mathcal{N}|$, be the set of stations where initial seismic waves have been recorded, and  $w_i(t)$ be the seismic wave of station $n_i \in \mathcal{N}^{\prime}$ at time $t$. Our goal is learn a function $F$ that can predict the seismic intensities, $I_1, I_2, ..., I_{N}$ of stations $n_1, n_2, ...n_{N}$, respectively, given the initial seismic waves of stations in $\mathcal{N}^{\prime}$. Mathematically, this can be formulated as $F : (w_1(t), w_2(t), ..., w_{|\mathcal{N}^{\prime}|}(t)) \mapsto (I_1, I_2, ..., I_{N})$.

Figure \ref{Probfor} shows a prototypical example of our problem setting. In Figure \ref{Probfor}(a), the epicenter of the earthquake is shown using a starlike symbol, and seven stations, $n_1, n_2, ...n_7$  are shown using circles,  where two stations, $n_1$ and $n_2$ marked as blue, have received initial seismic waves at time $t$, and the remaining five stations are yet to receive any signal. Now, we are to learn a function, $F$, that can predict the seismic intensity of all seven stations, as depicted in Figure \ref{Probfor}(b). The polygon heatmaps visually represent the intensity in this context.

\section{Background on Contrastive Learning and GNN}
This section provides a brief background on the key concepts of contrastive learning and graph neural networks (GNNs), which form the basis of our proposed method for predicting seismic intensity.

\subsection{Contrastive Learning} 
Contrastive learning is a self-supervised learning approach that aims to learn useful representations from unlabeled data by solving a pretext task \cite{oord2018representation}. The core idea behind contrastive learning is to encourage similarity between representations of similar or related instances while maximizing the dissimilarity between representations of dissimilar or unrelated instances. This is achieved by designing a contrastive loss function that minimizes the distance between positive pairs (related instances) and maximizes the distance between negative pairs (unrelated instances) in a latent representation space \cite{hadsell2006dimensionality}. 

Contrastive Learning begins with augmentation (see section IV.A for details) to create two viewpoints of a batch of input samples. When originating from the same sample, these views are regarded as positive samples; when from distinct samples, they are negative samples. An encoder network encodes, and a projection network maps the augmented samples to a feature space \cite{murshed2024self}. This feature space applies a well-designed contrastive loss function. Contrastive loss groups positive samples together and separates negative samples. Doing so clusters positive samples and disperses negative samples in the feature space \cite{contrastive_learning_method}.

\subsection{Graph Neural Networks (GNNs)}
Graph neural networks (GNNs) are a class of deep learning models designed specifically for learning from graph-structured data \cite{bronstein2017geometric,battaglia2018relational}. GNNs can effectively capture the complex relationships between nodes in a graph by iteratively aggregating and updating the node features through message-passing mechanisms \cite{gilmer2017neural}. GNNs can handle irregular and non-Euclidean data structures, making them well-suited for a wide range of applications, such as social network analysis \cite{berg2017graph}, molecular modeling \cite{duvenaud2015convolutional}, and geospatial analysis \cite{fan2022gnn}. 

The adjacency matrix is an essential element of GNNs, representing the graph's structure. The adjacency matrix, typically denoted as \textbf{A}, is a square matrix where each element $A_{ij}$ indicates the presence (often with a 1) or absence (with a 0) of an edge between nodes $i$ and $j$. This matrix is crucial for GNNs as it provides the necessary information about the interconnections between nodes in the graph, allowing the network to understand and learn from the graph's topological features. In some GNN variants, the adjacency matrix is enriched with edge weights or additional features that represent specific characteristics of connections between nodes.

Over time, GNNs have evolved into two primary categories: spectral methods and spatial methods. Spectral methods primarily concentrate on the eigenvalues and eigenvectors of a graph, whereas spatial methods prioritize the graph's connectivity \cite{gnn_two_types}. 

Graph Neural Networks (GNNs) are driven by a transfer function that updates the state vector of each node, incorporating neighborhood information into the node representation. The expression of the transfer function h\textsubscript{u} for spatial GNNs is as follows:

\begin{equation}
    h\textsubscript{u} = f(x\textsubscript{u}, x^e\textsubscript{u, ne[u]}, h\textsubscript{ne[u]}, x\textsubscript{ne[u]})
\end{equation}

where x\textsubscript{u} denotes the feature of the node, ${x^e}_{u, ne[u]}$ denotes the features of the edges connecting with u, h\textsubscript{ne[u]} denotes the transfer functions of the neighboring nodes of u, and x\textsubscript{ne[u]} denotes the features of the neighboring nodes of u. The expression of the output functions is as follows:

\begin{equation}
    o\textsubscript{u} = g(h\textsubscript{u}, x\textsubscript{u})
\end{equation}

The local transfer functions and the local output functions are applied to every node in the graph. By repeatedly iterating this process, the GNN gradually converges to a stable state \cite{gnn_eqn}. GNNs not only have shown efficacy in seismic intensity prediction but also in earthquake location and phase association, illustrating their versatility in seismology \cite{mousavi2023machine}. This is achieved by capturing spatial relationships between seismic stations and wave propagation patterns, leading to enhanced accuracy and generalization to unseen data. However, the application of GNNs across different seismic networks, such as adapting a model trained on California's network for use in Japan, presents challenges due to varying network configurations. This limitation underscores the need for cautious adaptation and potential model retraining for different geological settings.

\section{Proposed SC-GNN Methodology}

\begin{figure*}[! h]
\centering
    \includegraphics[width=1.7\columnwidth,height=70mm]{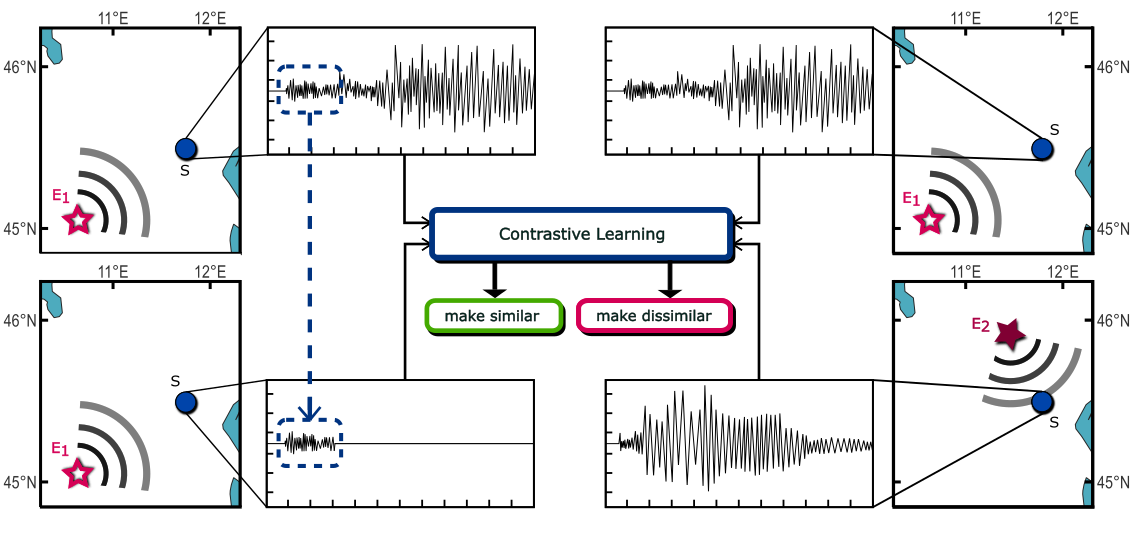}
    \caption{Contrastive learning within SC-GNN aligns seismic waveform representations from multiple stations, enhancing event-specific intensity prediction.}
    \vspace{-2mm}
    \label{Contrastive_intuition}
\end{figure*}

In this section, we propose a novel architecture, namely the seismic Contrastive Graph Neural Network (SC-GNN) model, that seamlessly integrates contrastive learning\cite{oord2018representation} with graph neural networks (GNN)\cite{battaglia2018relational} for the \emph{fast} and \emph{accurate} seismic intensity prediction. We use GNN to effectively capture spatiotemporal features from seismic waveform data from different stations spread across a geographical region. Despite their strengths, GNNs alone prove insufficient for early earthquake intensity prediction, as they rely on extended time window waveforms for feature extraction, which hinders real-time applications. Therefore, this paper introduces a contrastive learning-based approach to learning unique seismic embeddings for each earthquake event, even from shortened time-window waveform inputs, which ultimately help us in fast and accurate seismic intensity prediction.

\subsection{The Key Ideas of Contrastive Learning in Early Earthquake Prediction}
\label{subsec:contra}

Our contrastive approach primarily aims to learn unique seismic embeddings for each earthquake event, even from shortened time-window inputs. This approach is driven by the fundamental intuition of ensuring similarity in embeddings generated from differing input window lengths of the same earthquake event while maintaining distinctness in embeddings for separate events. Generating analogous embeddings from full and their corresponding truncated time windows implies that the model is capable of discerning the same features from reduced time windows as it does from extended ones. This proficiency is paramount for early warning systems, where precise intensity prediction from the shortest possible time window is the core of the task. Consequently, the sooner the model can make an accurate prediction, the earlier a warning can be issued, enhancing the efficacy and utility of the system.

Figure~\ref{Contrastive_intuition} visually illustrates the above concept. An original seismic waveform from earthquake $E_1$ is represented alongside a truncated, zero-padded augmentation, reflecting a shortened variant of the waveform. A distinct waveform is also depicted for event $E_2$. Our contrastive learning framework operates by promoting similarities between the original and augmented $E_1$ samples' embeddings while distinguishing the $E_1$ and $E_2$ samples' embeddings. This approach facilitates learning unique seismic embeddings by drawing parallels with the concept of positive and negative samples in a typical contrastive learning framework. Furthermore, as earthquake events do not require any pre-labelling for this task, our approach can be considered a self-supervised learning technique. This training structure enables our model to predict seismic intensities from significantly shorter time windows, empowering the model to deliver distinct intensity predictions for each individual earthquake.

\subsection{Architectural Overview of SC-GNN}
\label{subsec:arch}

\begin{figure*}[! h]
\centering
    \includegraphics[width=1.9\columnwidth,height=75mm]{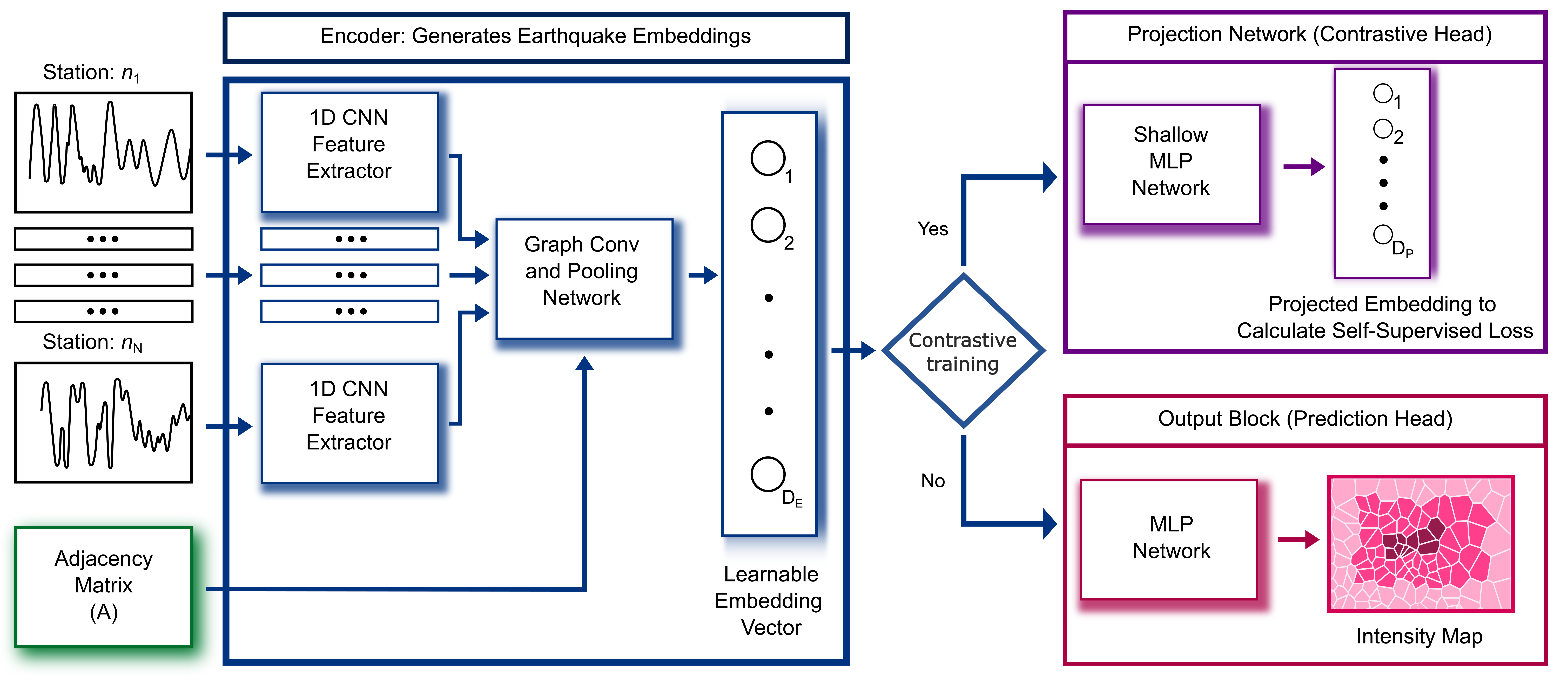}
    \caption{SC-GNN's workflow: Seismic data from stations feeds into a 1D CNN, then a GCN, followed by either contrastive learning or intensity prediction.}
    \vspace{-2mm}
    \label{SC-GNN_Functional}
\end{figure*}

A schematic representation of SC-GNN in terms of its functional blocks is shown in Fig. \ref{SC-GNN_Functional}.

\subsubsection{Input Block} 
The first point of interaction with the model begins with the input block. The SC-GNN model accepts 3-component seismic waveforms and an adjacency matrix representing the relationships between different seismic stations. The adjacency matrix is crucial in conveying spatial correlations between stations to the subsequent GNN block. Inputs are first subjected to TimeDistributed Batch Normalization, which not only optimizes the input scale for the network but also retains the relative amplitude information via the learned scale ($\gamma$) and shift ($\beta$) parameters, thereby ensuring the model's efficacy in intensity prediction tasks. We denote the 3-component waveform input of the i-th station by $w_i(t)$.

\textbf{Adjacency Matrix Preparation}: To adequately represent the graph structure of the seismic network, we prepare the adjacency matrix in a way that considers the mutual distances between stations, carefully considering the distance reciprocity and selective thresholding. Here, we outline the steps involved in preparing the adjacency matrix.

In the first step, we begin with an initial adjacency matrix which contains the pairwise distances between the seismic stations. However, considering our objective of inferring the intensity at distant stations, it is more intuitive to model the connections between the stations as a function of proximity rather than the actual distance. Hence, the entries of the adjacency matrix are inversed, thereby encoding the notion of reciprocal distance into the adjacency matrix. This results in closer stations having a higher edge weight than distant ones.

To ensure that the diagonal entries (representing self-loops) do not introduce bias into the network, they are initially filled with a large number, essentially representing an 'infinite' distance. After taking the reciprocal, these diagonal entries become infinitesimally small. To rectify this, the diagonal elements are replaced with the maximum value in the transformed adjacency matrix.

In the next step, the entire adjacency matrix is normalized by the maximum value, thereby ensuring that all edge weights fall within the range [0,1].

The final stage of preparing the adjacency matrix involves thresholding. We calculate the adjacency matrix's 75th percentile (the threshold for the top 25\% of values). All edge weights falling below this threshold are set to zero. This is done to sparsify the adjacency matrix, effectively retaining only the strongest connections in the graph and making the GNN computation more efficient. The threshold of the top 25\% has been determined through extensive empirical study. Algorithm \ref{alg:adjacency} succinctly presents this entire adjacency matrix preparation process.

This methodical process of adjacency matrix preparation enables the efficient representation of the seismic network, thereby ensuring that our GNN model accurately captures the crucial characteristics necessary for successful seismic intensity prediction.

\begin{algorithm}
\caption{Preparation of the adjacency matrix}\label{alg:adjacency}
\begin{algorithmic}[1]
\State Load matrix containing mutual inter-station distances in Km
\State Fill the diagonal of the matrix with a large value
\State Compute reciprocal of matrix elements, scale by minimum value in the matrix
\State Replace diagonal elements with the maximum value in the matrix
\State Normalize the matrix by its maximum value
\State Compute the threshold as the 75th percentile of matrix elements
\State Set all elements below the threshold to zero
\end{algorithmic}
\end{algorithm}

\subsubsection{1D CNN Feature Extractor}
The normalized seismic waveforms are then processed by a series of TimeDistributed 1D Convolutional (Conv1D) and MaxPooling layers. These layers, equipped with \emph{swish} activation functions, capture and highlight the essential characteristics in the waveform data. Dropout layers are added to promote model robustness and mitigate overfitting, providing a form of regularization. Following this, the extracted features undergo further refinement via multiple Conv1D layers with varying kernel sizes and filters as shown in Fig. \ref{SC-GNN_Functional} (1D CNN Feature Extractor). Additional Dropout and Batch Normalization layers are interspersed for enhanced regularization and normalization of the feature vectors. Finally, the extracted features are flattened and passed through multiple dense layers to reduce the length of the feature vectors. Mathematically, we can represent the operation of the 1D CNN feature extractor by
\begin{equation}
    F_i = f_{CNN}(w_i(t));\;\;\;\; 
    \forall i \in \{1,2,...., N\},
\end{equation}
where $F_i$ denotes the extracted features for the seismic waveforms from the $i$-th station and $f_{CNN}(.)$ indicates the feature extraction operation performed by the 1D CNN block of SC-GNN.

In terms of explicit construction, this 1D CNN feature extractor consists of three Conv1D blocks, one after another. Each block consists of two Conv1D layers, followed by a Batchnorm, a max-pooling and a drop-out layer. As we traverse deeper into the architecture, the number of filters in the Conv1D layers progressively increases while kernel sizes decrease. The initial Conv1D layer applies 16 filters with a kernel size of 50, whereas the deepest layer uses 96 filters with a kernel size of 4. Moreover, Max Pooling layers have pool sizes that progressively increase (4, 6, and 12).

The first convolutional block applies zero padding to prevent excessive feature dimension reduction, while the last two blocks avoid it to limit noise insertion. After the Conv1D blocks, the output is flattened and processed through two dense layers of 256 neurons each, reducing the dimensionality for the GNN block, where they serve as node features. A Dropout layer is included post-dense layers to counteract overfitting and better generalization \cite{murshed2021deep}.

\subsubsection{Graph Neural Network Block}
After the feature extraction, the data advances to the GNN block. In this segment, the temporal information from the seismic waveforms is paired with the spatial relationships embedded in the adjacency matrix. A couple of ChebConv layers having 256 channels and Dropout layers are applied to the waveform data, followed by a GCSConv layer with 256 channels. This approach enables the model to encode the complex spatiotemporal relationships present in the seismic data, an advantage over conventional Convolutional Neural Networks that might miss such intricate correlations. We express the operation of the GNN block mathematically by
\begin{equation}
    G_i = f_{GNN}(F_i, A);\;\;\;\; 
    \forall i \in \{1,2,...., N\},
\end{equation}
where $G_i$ denotes the output of the GNN block corresponding to the input CNN feature vector $F_i$, the adjacency matrix $A$, and the graph convolutional and pooling operations, $f_{GNN}(.)$, performed by the GNN block of SC-GNN.

To gain a deeper understanding of the GNN block in our proposed SC-GNN model, we need to elaborate on the distinct GNN layers utilized - ChebConv (Chebyshev Convolution) and GCSConv (Graph Skip Convolution).

\begin{itemize}
    \item ChebConv:  The ChebConv layer in a GNN employs Chebyshev polynomials to approximate the spectral filter derived from GNN \cite{sahbi2021learning}. This facilitates the execution of convolution operations over the graph structure, thereby encapsulating both local and non-local information.
    \item GCSConv: The GCSConv layer, an evolved variant of the GCN layer with trainable skip connections, enhances the local structure and feature diffusion \cite{li2020deepergcn}. It accomplishes this by amalgamating spectral diffusion, which focuses on the graph's global structure, akin to ChebConv, and spatial diffusion, which emphasizes each node's local neighbourhood structure.
\end{itemize}

The intuition behind using two ChebConv layers followed by GCSConv is to initially capture global graph features via the ChebConv layers, subsequently refining these features by considering local neighborhood information using the GCSConv layer. This strategy is designed to be especially beneficial in regression tasks such as intensity prediction, where precision is paramount. It potentially harnesses both the global and local graph structures to generate more accurate predictions. The dual usage of ChebConv and GCSConv layers fosters a more comprehensive representation of the graph as they may capture disparate features.

\subsubsection{Embedding Layer}
The data exits the GNN block and enters the Embedding Layer, where a GlobalAttentionPool layer \cite{li2017gated} with five channels is used. This layer aids the model in focusing on the most informative parts of the graph concerning the prediction task \cite{li2015gated}. Along with the attention mechanism, the data passes another Dense layer having 100 neurons. Here, the model generates the final embeddings by concatenating the GlobalAttentionPool and Dense layers' outputs, encapsulating each earthquake's distinct seismic characteristics. Mathematically, we can represent this embedding generation as
\begin{equation}
    Emb_i = f_{emb}(G_i);\;\;\;\; 
    \forall i \in \{1,2,...., N\},
\end{equation}
where $Emb_i$ denotes the embeddings for the i-th station with length $D_E = 32$ and $f_{emb}(.)$ indicates the embedding generation operation performed by the embedding layer of SC-GNN.

\subsubsection{Contrastive Head}
In the final stage of the SC-GNN architecture, the Projection Network employs a multi-layer perceptron (MLP)  to project the seismic embeddings into a lower dimensional space. These embeddings essentially are feature representations of a snapshot of seismic wavefield across the seismic network. Using a series of Dense layers, supplemented with Dropout (rate $= 0.05$) and Batch Normalization for stability and regularization, the network projects the embeddings into a space conducive to the contrastive learning objective. Similar earthquakes are positioned close together in this space, and dissimilar ones are spread apart. The projection operation can be expressed as
\begin{equation}
    z_i = P(Emb_i);\;\;\;\; 
    \forall i \in \{1,2,...., N\},
\end{equation}
where $z_i$ denotes the embedding projection for the i-th station embedding with length $D_p = 10$ and $P(.)$ represents the projection operation executed by the projection head of SC-GNN. Now, the embedding project vector, $z$, which is used directly in the contrastive loss function, is given by
\begin{equation}
z = \begin{bmatrix} z_1 \\ z_2 \\ \vdots \\ z_N \end{bmatrix}.
\end{equation}

Note that our model architecture's contrastive head is disposable. Its main purpose is to generate projection embeddings for self-supervised contrastive loss calculations. As a result, it plays a crucial role in the hybrid contrastive training phase. Afterward, it is removed from the subsequent stages of implementation as contrastive loss calculations are not required for the goal of prediction. The self-supervised contrastive loss function \cite{khosla2020supervised}, $\mathcal{L}^{\text{cont}}$, applied at this stage is given by
\begin{equation}
\mathcal{L}^{\text{cont}} = \sum_{m \in \mathcal{M}}^{} \mathcal{L}_m^{\text{cont}} = - \sum_{m \in \mathcal{M}}^{} \log \frac{\exp(z_m \cdot z_{m'}/\tau)}{\sum\limits_{a \in A(m)}^{} \exp(z_m \cdot z_a/\tau)}.
\end{equation}
where, $\mathcal{M}= {\{1,2, \ldots M\}}$,  represents a training batch with all the odd samples being original samples and the adjacent even samples corresponding to their respective augmented samples, together forming a positive pair. For each sample $m \in \mathcal{M}$ and its corresponding positive sample $m' \in \mathcal{M}$, $A(m)= \mathcal{M} \setminus m$. $z$ denotes the $D_p$  dimensional projection of the $D_E$ dimensional representation, and $\tau$ is the temperature parameter, adjusting the sensitivity of the loss function to the similarity between samples. The augmentation we perform here is unique. Instead of the typical 1D signal augmentations such as noise injection, scaling, time-shifting, time-scaling, frequency shifting, etc., we clip and zero-pad the original signal to generate the augmented signal as shown in Figure \ref{Contrastive_intuition}. We elaborate more on this augmentation technique in the next sub-section (\ref{subsec:traninginf}).

In parallel with the contrastive loss, we also apply another regression loss to ensure that the generated embeddings are task-specific, i.e., specifically oriented toward accurate intensity prediction. While the contrastive loss is applied to the outputs of the contrastive head, this regression loss \cite{murshed2023cnn} is applied to the outputs of the prediction head. This regression loss is given by
\begin{equation}
\label{prop_loss}
\begin{split}
{\mathcal{L}^{\mathrm{reg}} = w_1\times(1-r^{2}) +w_2\times L_{\mathrm{HL}}  + w_3 \times \epsilon^{2} + w_4\times  |\epsilon| }\\
+ w_5\times L^a_{\mathrm{HL}},
\end{split}
\end{equation}
where $L_{\mathrm{HL}}$ denotes the well-known HL value, $L_{\mathrm{HL}}$ is a modified asymmetric version of the Huber-loss, $r$ represents the Pearson correlation coefficient, $\epsilon$ is the error in the prediction, and $w_1,w_2,w_3,w_4$ and $w_5$ are the weights applied to the correlation loss, HL, MSE, MAE and the asymmetric HL, respectively. The exact values used for $w_1,w_2,w_3,w_4$ and $w_5$ are determined to be $0.002, 1.0, 0.0096, 0.002$ and $0.0032$, respectively, through extensive trial and error to suit the required task best. The $\mathcal{L}^{\mathrm{reg}}$ function is a custom loss function that combines elements of Huber loss (HL) and correlation loss. The HL component is less sensitive to outliers in the data, providing a more stable training process. In contrast, the correlation component ensures the predicted seismic intensities are closely aligned with the actual values. Notably, none of the well-known regression losses (i.e., MSE, MAE, mean absolute percentage error, etc.) except HL solely performed as well in our experiments.

The $\mathcal{L}^{\mathrm{hyb}}$ function is a hybrid of the $\mathcal{L}^{\mathrm{cont}}$ and $\mathcal{L}^{\mathrm{reg}}$ functions, comprehensively evaluating the model's performance during the contrastive training stage.  It is a straightforward sum of $\mathcal{L}^{\mathrm{cont}}$ and $\mathcal{L}^{\mathrm{reg}}$, i.e, $\mathcal{L}^{\mathrm{hyb}} = \mathcal{L}^{\mathrm{cont}} + \mathcal{L}^{\mathrm{reg}}$.

Following the conclusion of the hybrid contrastive training phase, the contrastive head is removed as it serves no further purpose for the prediction stage. Specifically, the contrastive head is omitted during the regression training phase (fine-tuning) and subsequent inference tasks. This is due to the different requirements of these subsequent phases, which do not involve calculating contrastive loss and, therefore, do not need the projection embeddings generated by the contrastive head. Therefore, while the contrastive head is crucial for the initial self-supervised learning and embedding generation during the hybrid contrastive training phase, its functionality is deliberately limited to this stage, underscoring the disposable nature of this module within our model's architecture.

\subsubsection{Output Block (Prediction Head)}
After passing through the embedding layer, the processed data reaches the Output Block. In this block, the model generates the final predictions and embeddings. The SC-GNN model generates two forms of output: seismic intensity predictions at various seismic stations and seismic embeddings. The intensity predictions, produced by a sequence of Dense layers and a 'relu' activation function, represent the potential earthquake intensities at different seismic stations. On the other hand, the seismic embeddings are produced through another sequence of Dense layers, capturing the distinct characteristics of each earthquake.

This two-fold output serves multiple purposes. On the one hand, it aids in understanding the specifics of an earthquake's characteristics by inspecting the embeddings. On the other hand, it allows for predicting seismic intensities at various stations, a critical component for effective earthquake early warning systems. Mathematically, the seismic intensity predictions can be expressed as
\begin{equation}
    I_i = f_{out}(Emb_i);\;\;\;\; 
    \forall i \in \{1,2,...., N\},
\end{equation}
where $I_i$ denotes the seismic intensity for the i-th station and $f_{out}(.)$ represents the operation executed by the output block of SC-GNN. In the prediction head, we utilize the $\mathcal{L}^{\mathrm{reg}}$ loss function to train the generated intensity predictions. 

The SC-GNN architecture, with its dedicated functional blocks, successfully extracts, refines, and leverages seismic data to generate precise seismic intensity predictions. The individual blocks, each with its unique functionality, work together seamlessly, creating a robust and efficient model. This architecture provides an advanced solution for early warning generation in the realm of earthquake predictions.

\subsection{Training Process of SC-GNN}
\label{subsec:traninginf}
In this subsection, we provide a detailed overview of the training and inference procedures of the SC-GNN model. The SC-GNN model leverages a unique combination of contrastive training and regression to create robust, task-specific embeddings. The sub-section is divided into two segments: model training and inference.
\subsubsection{Model Training}

\begin{figure}[t]
    \includegraphics[width=\linewidth]{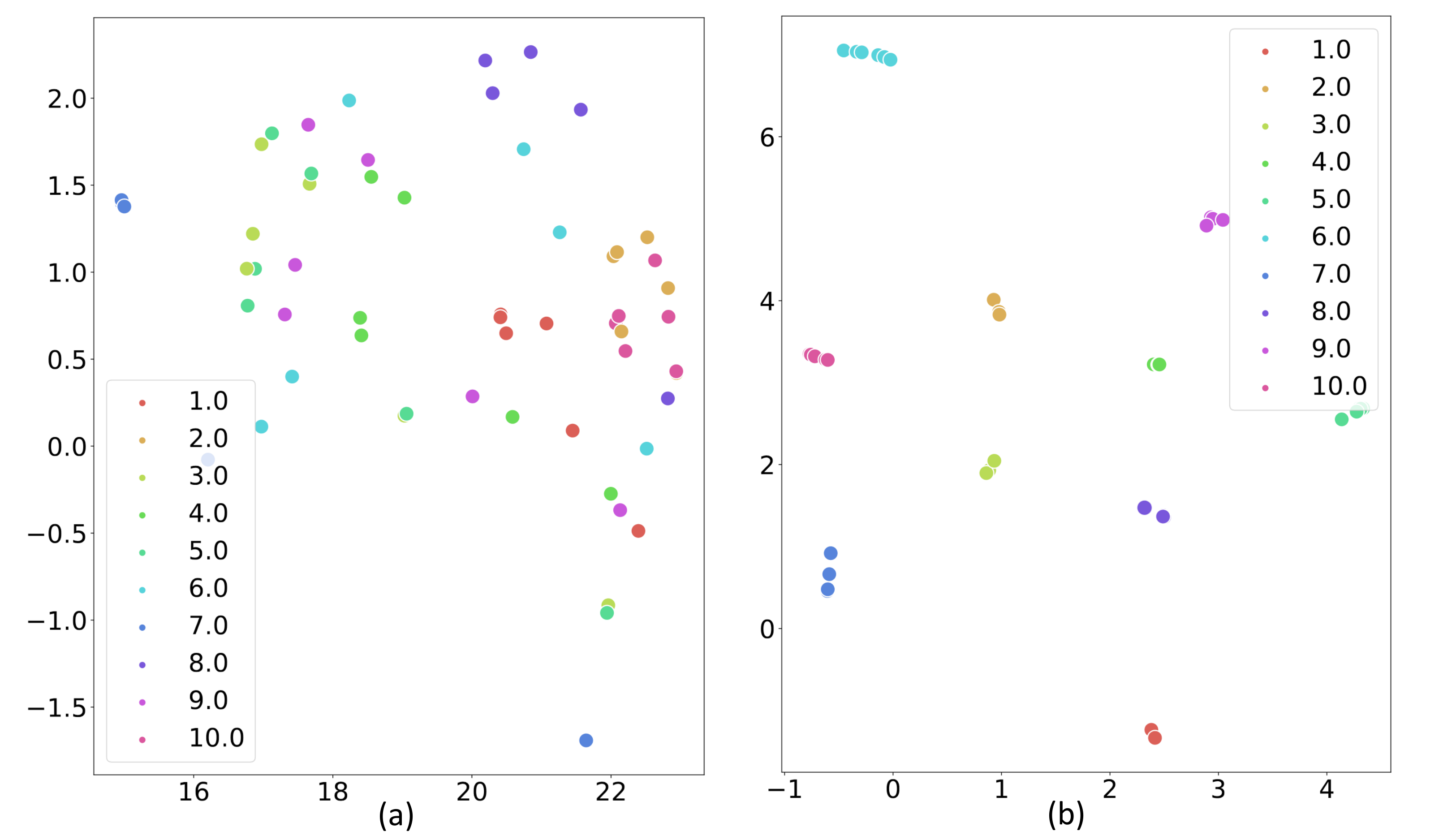}
    \caption{TSNE plot displaying seismic event embeddings: (a) before contrastive training (b) after contrastive training.}
    \vspace{-2mm}
    \label{Fig_TSNE}
\end{figure}

\begin{figure}[t]
    \includegraphics[width=\linewidth]{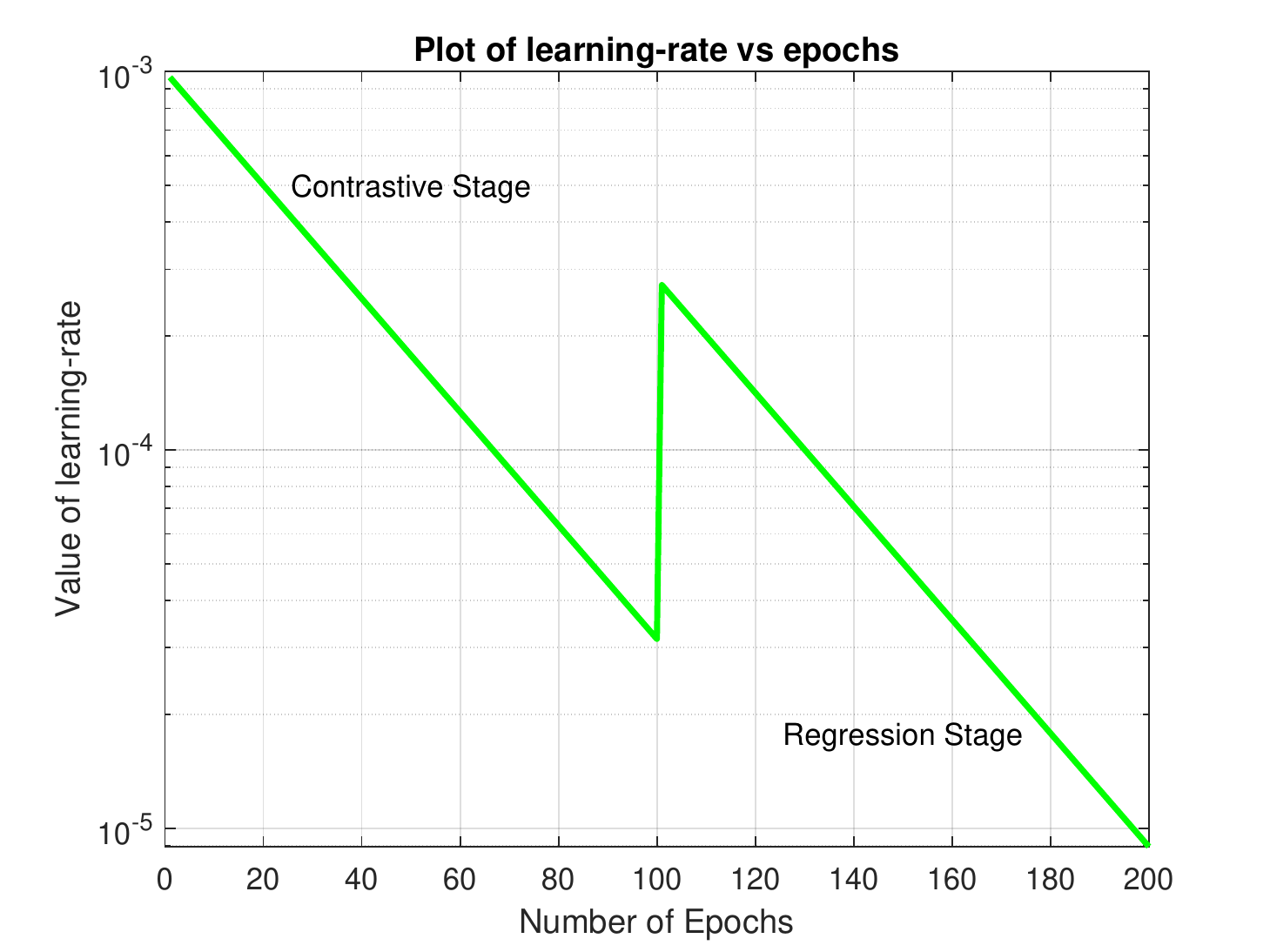}
    \caption{Learning-rate Scheduler used to train SC-GNN. }
    \vspace{-2mm}
    \label{lr_scheduler}
\end{figure}

The training process of the SC-GNN model unfolds in two primary stages. The initial phase employs contrastive training with a distinct data augmentation scheme. Each seismic waveform, denoted as $w_i(t)$, undergoes transformation to produce augmented samples, $w^a_i(t)$. The augmentation procedure involves clipping the original waveform to $t_c$ seconds, where $t_c$ is an integer uniformly selected from the set $\{5, 10, 15, 20, 25\}$, followed by zero-padding to maintain a consistent input length of 30 seconds\footnote{3000 samples at a fixed sampling frequency of 100 Hz}. This process yields five augmented samples per original waveform. Selecting distinct window-length inputs from the same seismic event forms positive pairs, while waveform data from disparate events generate negative pairs.

This contrastive learning phase employs a novel hybrid loss function, $\mathcal{L}^{\mathrm{hyb}}$, combining both contrastive and regression losses. This approach aims to direct the generation of task-specific embeddings for seismic intensity prediction. The model is trained under this hybrid loss function for the first 100 epochs. The seismic embeddings generated during the training phase are exemplified by the TSNE plot depicted in Fig. \ref{Fig_TSNE}. The plot showcases embeddings for ten distinct seismic events, labeled 1 through 10, alongside their respective augmented samples with varying input window lengths before (Fig. \ref{Fig_TSNE} (a)) and after the contrastive training stage (Fig. \ref{Fig_TSNE} (b)). Notably, the embeddings of augmented samples belonging to the same seismic event but with different window lengths (5, 10, 15, 20, 25, and 30s) are clustered together after the contrastive training phase, indicating that the model has successfully learned to generate similar embeddings for augmented samples from the same event. Furthermore, distinct seismic events form separate unique groups, illustrating the model's ability to differentiate between individual events. It is important to mention that each distinct earthquake is represented by a unique color in Fig. \ref{Fig_TSNE}, while augmented samples belonging to the same event are depicted in the same color.

The second training phase centers on fine-tuning the model for seismic intensity prediction at various geographical locations using the input earthquake data. This is approached as a regression task, with the majority of the model layers being frozen, preserving the integrity of the embeddings learned during the contrastive training phase. The regression training is guided solely by the regression loss function $\mathcal{L}^{\mathrm{reg}}$, and it continues for an additional 100 epochs.

The training process employs the Adam optimizer, owing to its efficient handling of large-scale data. An exponential decay learning rate scheduler, as illustrated in Fig. \ref{lr_scheduler}, assists in stabilizing the training process over a total of 200 epochs. The batch size is set at 32, balancing adequate learning and manageable computational requirements. The model's best-performing weights during the training process are captured using checkpoints to ensure that the optimal model configuration is preserved. Algorithm \ref{alg:training} concisely presents the SC-GNN training steps. Like every DL-based model, our method's generalizability depends on the training data. Retraining the model is not needed if the test dataset includes a few new stations adjacent to the training data. For a test set from a different region, the model would benefit from transfer learning or a fresh training cycle.

\begin{algorithm}
\caption{Training Process of the SC-GNN}\label{alg:training}
\begin{algorithmic}[1] 
\Require Training data: Waveforms, Inter-station Distances, Shakemap Labels of Intensities
\Ensure Predicted intensities at all stations
\State Convert all PGA values to intensity using the EMS conversion equation using \ref{Eq_EMS}
\State Impute all missing waveforms with zeroes
\State Generate augmented samples by clipping and zero-padding the waveforms
\State Prepare the adjacency matrix using inter-station distances
\State Segment the data into training, validation, and test sets for cross-fold validation
\State Train the SC-GNN model for 100 epochs using $\mathcal{L}_{\mathrm{hyb}}$
\State Freeze all layers up to the embedding layer and discard the contrastive head
\State Load the best model weights based on the validation metric, $\mathcal{M}_{\mathrm{val}} = \mathcal{L}_{\mathrm{val}}^{\mathrm{cont}} + 100 \times \mathcal{L}_{\mathrm{val}}^{\mathrm{reg}}$
\State Train the model for another 100 epochs using $\mathcal{L}_{\mathrm{reg}}$
\State Load the best model weights based on the validation metric, which is the same as the validation loss $\mathcal{L}_{\mathrm{val}}^{\mathrm{reg}}$
\State Pass the test data for inference, i.e., intensity prediction
\end{algorithmic}
\end{algorithm}

\subsubsection{Model Inference}
\label{subsec:infer}
Upon completing the model training, we progress to the inference stage. Any window length of input seismic waveforms for model inference is accepted, provided they are zero-padded to comply with the fixed input size. Additionally, the corresponding adjacency matrix must also be inputted. The model outputs seismic intensities at all geographic points contained within the adjacency matrix. This functionality empowers the model to generate real-time, reliable seismic intensity predictions.

Along with the intensity predictions, the SC-GNN also produces the embedding vector, $Emb$, that represents the entire seismic event or earthquake, and it is given by
\begin{equation}
Emb = \begin{bmatrix} Emb_1 \\ Emb_2 \\ \vdots \\ Emb_N \end{bmatrix}.
\end{equation}

Without the loss of generality, though the above embedding vector is applied for our intensity prediction task, it can potentially be used to predict various other seismic parameters.

\section{Experimental Evaluation}

In this section, we assess the performance of our proposed SC-GNN model for real-time seismic intensity prediction using three real-world seismic datasets. Additionally, we compare the effectiveness of our proposed model against several state-of-the-art baseline models by examining standard performance metrics.

\subsection{Environment}

The proposed model is implemented using Python, Tensorflow 2.12, and Keras 2.12. The GNN layers are imported from Spektral 1.2.0. Numpy is utilized for calculations, and some figures are generated using MATLAB 2022b. The model is trained on Google Colab Pro+ with 83.5 GB system RAM, 166.8 GB disk space, and an NVIDIA A100-SXM4-40GB GPU to expedite the training process. Following training, the model's size is compact (approximately 2.8 MB), allowing deployment on standard PC hardware and edge computing platforms. Notably, the baseline model simulations are conducted using the same configuration for a fair comparison.

\subsection{Data Description}

We use the following three widely used datasets to demonstrate the prediction performance of our proposed algorithm.

\subsubsection{Central Italy (CI) Dataset}

As delineated in \cite{jozinovi_2020_cnnpredim_rapid}, this dataset draws upon three-component waveforms from 915 earthquake events captured by an extensive network of 39 stations across Central Italy. It comprises three distinct channels of waveforms: HN, HH, and EH. The EH channels harbor waveforms at a 100 Hz or 125 Hz sampling rate, while the HH and HN channels exclusively accommodate 100 Hz waveforms. However, all the waveforms are resampled to a fixed uniform sampling rate of 100 Hz in the finalized dataset.

To employ this dataset for early warning analysis, it is essential to identify the p-waves using Phasenet \cite{zhu_2018_phasenet}. Phasenet is designed to process time-series seismic data and categorize them into P-wave, S-wave, and noise. This necessitates the modification of waveform sampling rates that deviate from 100 Hz, ensuring a uniform rate throughout. Furthermore, Phasenet requires waveforms of equal length. After discerning the longest waveform length to be 10310, shorter waveforms are extended to match this length via zero padding. Based on two criteria—capturing time of p-picks and phase score—Phasenet makes p-wave pick predictions. 

The input dataset is generated by extracting the first 30 seconds of all waveforms, starting from the respective earthquake origin times. For earthquakes with magnitudes (M) below 4, HH and EH channels are employed, while HN channels are utilized for earthquakes with M greater than or equal to 4. The ground truth is generated using the peak ground acceleration (PGA), which is then converted to intensity through the ground motion-to-intensity conversion equation (GMICE) described later. Instances, where waveform data for specific earthquakes are absent from some stations, are addressed by zero-filling. The ground truth in those cases is imputed by utilizing USGS shakemap version 4.1 \cite{shakemap}.

Due to the availability of a higher number of seismic events and waveforms, we choose this CI dataset as our primary dataset. Unless otherwise mentioned, the experimental results presented in the subsequent discussions are generated using this dataset.

\subsubsection{Central Western Italy (CW) dataset}
This dataset consists of 3-component waveforms of 266 earthquakes from central western Italy recorded by 39 stations. It is thoroughly described in \cite{central_western_italy_dataset}, and the prepared dataset can be found in \cite{CWI_prepared_dataset}. Similar to the CI dataset, intensity labels are generated by converting the PGA values to intensity using the GMICE. It is important to note that all the waveforms available for the CW dataset are of length 10s. Hence, the value of $t_c$ for augmented sample generation during the contrastive training is confined to the set $\{5, 6, 7, 8, 9\}$, and the input length is 1000 samples at a fixed sampling frequency of 100 Hz.

\subsubsection{STEAD (STanford EArthquake Dataset)}
The STEAD ~\cite{stead,murshed2023analysis} serves as our third dataset for this study. Specifically, we focus on the region of \emph{California}, limited within the geographical coordinates of latitude 32.5\textdegree\ to 35\textdegree\ and longitude -115\textdegree\ to -118\textdegree, as this area encompasses the majority of earthquakes documented in the dataset. Within California, we identified 191 earthquakes that a union of 194 seismic stations recorded. Each station's recordings consist of 3-component seismograms, with each component spanning 1 minute. The instrument responses associated with the stations, obtained from the Incorporated Research Institutions for Seismology (IRIS) \cite{2019_iris}, were acquired and converted into acceleration units \cite{stead}. However, we extract the initial 30 seconds from each waveform to generate the final input dataset, following the same approach applied to the CI dataset.

To impute the missing PGA values in stations where waveform data are unavailable, USGS shakemap version 4.1 \cite{shakemap} is used to make PGA prediction with the help of instructions from shakemap manual \cite{shakemap_manual}. Using Shakemap, we generate XML grids of interpolated ground shaking data for each earthquake. XML grid consists of thousands of location-specific data fields around the earthquake's epicenter spaced at 0.0055\textdegree\ $\sim$ 0.0167\textdegree\ intervals of latitude and longitude. Each field contains PGA, PGV, MMI, and other attributes of a particular location. The process of installation and grid file generation using Shakemap is described in videos given in the online repository. The PGA values of the 194 stations are obtained from the field having the nearest latitude and longitude from each station. Again, we convert the PGA values to intensity using GMICE.

The key attributes of the datasets are shown in Table \ref{dataset_comparison}.

\begin{table}[h]
\centering
\begin{tabular}{|p{1.5cm}|p{1.5cm}|p{1.5cm}|p{1.5cm}|}
\hline
\textbf{Properties} & \textbf{CI} & \textbf{CW} & \textbf{STEAD} \emph{(California)} \\ \hline
Earthquakes & 915 & 266 & 191 \\ \hline
Stations & 39 & 39 & 194 \\ \hline
Available Waveforms & 33175 & 9856 & 4299 \\ \hline
Magnitudes (M) & $2.9 \leq M \leq 6.5$ & $2.9 \leq M \leq 5.1$ & $1.7 \leq M \leq 5.42$ \\ \hline
Mean of M & 3.32 & 3.29 & 2.91 \\ \hline
PGA ($cm/s^2$) & $0 \leq M \leq 0.49$ & $0 \leq M \leq 0.029$ & $0 \leq M \leq 0.015$ \\ \hline
PGA Mean($cm/s^2$) & 0.013 & 0.0014 & 0.0019 \\ \hline
Periods & 1-1-16 to 29-11-16 & 1-1-13 to 20-11-17 & 19-2-10 to 9-4-18 \\ \hline
\end{tabular}
\caption{Comparison of the Datasets}
\label{dataset_comparison}
\end{table}

\subsection{Ground Motion to Intensity Conversion}
The implications of an earthquake at a specific site demand the determination of intensity value at that locale. The previously employed Ground Motion Prediction Equations (GMPEs) yield PGA values, necessitating a conversion into intensity. For this transformation, we utilize the Ground Motion to Intensity Conversion Equation (GMICE) proposed by Zanini \textit{et al.} \cite{zanini2019}.

This equation generates intensity values following the European Macroseismic Scale (EMS-98), originating from regression analysis of Italian seismic data collected from the Parametric Catalogue of Italian Earthquakes and ITACA. The equation is as follows:

\begin{equation}\label{Eq_EMS}
I_{\text{EMS-98}} = 2.03 + 2.28 \log(PGA(cm/s^2)) .
\end{equation}

It is important to observe that Equation \ref{Eq_EMS} was specifically designed to operate within the range $2 \leq I_{\text{EMS-98}} \leq 9.5$. Consequently, any values resulting from this equation are confined within this predefined range to ensure validity. Specifically, if a conversion using this equation results in an $I_{\text{EMS-98}}$ value less than 2, the value is elevated to meet the minimum threshold of 2. This procedure is conducted to adhere to the stipulated constraints of the GMICE described by Equation \ref{Eq_EMS}.

\subsection{Baseline Models}

To demonstrate the effectiveness of our proposed approach, we have compared the performance of our proposed SC-GNN with some state-of-the-art models for seismic intensity prediction. These models, along with their adopted parameter settings, are described below:

\subsubsection{GMPE \cite{bindi2011,boore2014}}
The first baseline model is regression-based ground motion prediction equations (GMPE), considering the latest release of the strong motion database. These predict PGA, peak ground velocity (PGV), and 5\%-damped spectral acceleration over a magnitude range of 4-6.9 and distances up to 200 km. The total standard deviation confirms the large variability of ground shaking parameters for regional datasets containing small to moderate-magnitude events. This model is an update of the ITA08 GMPE \cite{bindi2010horizontal}, considering improved data, reprocessing, and an extended distance range. 

For comparison purposes with our proposed SC-GNN, in the case of the CI and CW dataset, the prediction of PGA values is achieved via the GMPE, as formulated by Bindi \textit{et al.} \cite{bindi2011}. This equation, deriving from the Italian strong motion database, "Italian Accelerometric Archive" (ITACA), necessitates magnitude, station distance, and soil site class as inputs. Ideally, the GMPE utilizes the Joyner-Boore distance (RJB); however, the lack of fault geometric details in this instance compels us to substitute it with the epicentral distance—a deviation acceptable within the purview of this equation. Identification of local site classes at respective station locations is carried out using the standalone software Soil Class-Italy (SSC-Italy). This software applies the Eurocode 8 soil classification developed by Forte \textit{et al.} \cite{forte2019}.

For STEAD, PGA values at the station locations for each earthquake are predicted using GMPE given by Boore \textit{et al.} \cite{boore2014}. This GMPE is developed using the NGA-West2 ground motion database provided by the Pacific Earthquake Engineering Research Center (PEER). This equation requires magnitude, distance of the site, and local site effects in terms of shear wave velocity (Vs30). Similar to the CI and CW datasets, we have used epicentral distance instead of Joyner–Boore distance (RJB). For identifying the Vs30 value at the location of the stations, we have used the Global Vs30 grid file, which is made available by the United States Geological Survey (USGS) based on the works
of Heath \textit{et al.} \cite{heath2020}. The final location coordinates and magnitudes, as documented in the datasets, were employed for applying the GMPEs to all three datasets.

\subsubsection{CNN Based Model \cite{ground_motion_intensity}}
The second baseline model is a deep Convolutional Neural Network (CNN) designed to predict earthquake ground shaking intensity measurements using multistation 3C acceleration waveforms. The input data consists of normalized waveform data from various seismic stations, and the model does not require prior knowledge of the earthquake source. The CNN architecture is adapted from Kriegerowski \textit{et al.} \cite{kriegerowski2019deep} and consists of three convolutional layers followed by fully connected layers. The first two convolutional layers learn temporal patterns station-by-station, while the third layer gathers cross-station information. The model has been tested on raw data without data pre-selection and has shown stability and accurate prediction of ground shaking intensity. The technique is not designed for earthquake early warning but provides useful estimates of ground motions within 15-20 seconds after the earthquake origin time.

\subsubsection{GCN Based Model \cite{bloemheuvel2022graph}}
Recently, a Graph Convolutional Network (GCN) based approach, named TISER-GCN, has been proposed for multi-variate time-series regression that achieves state-of-the-art performance. It predicts ground-shaking intensity at seismic stations using a regression approach. The model utilizes two 1D CNN layers with wide kernel sizes, small strides, increasing filters, and ReLU activation functions to learn the temporal patterns of each station. After feature extraction, the model combines node features (latitude, longitude) with partially flattened feature vectors to create input for the Graph Convolutional Network (GCN) layers. The GCN layers reduce the dimensions of the input and handle cross-station information. In contrast to standard graph pooling techniques, this model uses a flattened output from the final GCN layer to preserve meaningful features. The architecture concludes with a dense layer and five linear activations function-based dense layers that predict target variables such as PGA, PGV, etc. Here, we only utilize the PGA output to calculate the seismic intensity using the GMICE.

\subsection{Performance Metrics}

We have used a range of performance metrics to evaluate and compare the effectiveness of the proposed model with the baseline models.

    \textbf{Mean Squared Error (MSE):} MSE measures the average squared difference between the predicted and actual values. It is widely used in regression problems to quantify the error in predictions. A lower MSE indicates better model performance, with zero being the ideal value.

    \textbf{Standard Deviation (SD):} The standard deviation (SD) of the error represents the residuals' dispersion around the mean. It helps in understanding the variability of the model's predictions. A lower SD indicates that the model's predictions are more consistent and reliable.

    \textbf{Correlation Coefficient and $R^2$:} The correlation coefficient (CC) measures the strength and direction of the relationship between the predicted and actual values. A value close to 1 ($100\%$) indicates a strong positive correlation, while a value close to 0 represents no correlation at all. A high CC signifies that the model's predictions align with the actual values. The $R^2$ or the Coefficient of Determination is the square of the CC that measures how well the model's predictions explain the variance in the actual values.

Additionally, conditional scatter plots and Bland-Altman plots are used to assess the models' performance, biases, and generalization capabilities. By using these metrics, we thoroughly assess the effectiveness of the proposed model and compare it with the baseline models, considering various aspects such as prediction accuracy, consistency, and reliability.

\subsection{Comparison with Baseline Models}

\begin{table}[h]
\centering
\begin{tabular}{|c|c|c|c|c|}
\hline
Metric/Model & \textbf{SC-GNN} & TISER-GCN & CNN & GMPE \\ \hline
MSE          & 0.4172       & 0.9645 & 1.4027 & 1.3507 \\ \hline
SD           & 0.6110       & 0.9005 & 0.9701 & 1.0979 \\ \hline
CC & 83.94\%  & 61.34\% & 48.42\% & 43.11\% \\ \hline
\( R^2 \) & 70.46\% & 37.63\% & 23.44\% & 18.58\% \\ \hline
\end{tabular}
\caption{Comparison of the proposed SC-GNN model with baseline models on the CI dataset for $2 < I_{\text{EMS-98}} < 9.5$.}
\label{table:ci-dataset}
\end{table}

\subsubsection{Performance Comparison with Baselines}

In this section, we compare the performance of the proposed SC-GNN model with the baseline models: TISER-GCN, CNN, and GMPE, on our primary dataset with a 10s input window. The results are presented in Table \ref{table:ci-dataset}.

The proposed model, SC-GNN, outperforms the baseline models across all metrics in the CI dataset. Our SC-GNN model achieves the lowest MSE of 0.4172, which reflects around \emph{234\%} improvement over the state-of-the-art best-forming TISER-GCN model, indicating more accurate predictions. The SD of the error for the GNN model is the lowest at 0.6111, suggesting more consistent and reliable predictions than the other models. Furthermore, the SC-GNN model has the highest CC of 83.94\%, signifying a strong positive relationship between the predicted and actual values.

The significant improvement in performance metrics for our SC-GNN model can be attributed to its ability to capture intricate spatial and temporal patterns inherent in earthquake data by utilizing sophisticated GNN layers. Specifically, the ChebConv and GCSConv layers integrated within the SC-GNN effectively capture both local and non-local information encoded within the seismic graph, allowing the model to better understand the underlying structural dynamics of the data. In addition, the seismic embeddings generated during the self-supervised contrastive training phase acquire key traits ingrained in the extended seismic waveforms. This leads to more accurate and reliable predictions. In contrast, the baseline models may struggle to capture these relationships due to their respective limitations in handling the data's spatial and temporal aspects and lack of any contrastive learning phase.

\subsubsection{The Effect of Varying Time Window}

In this section, we compare the performance of the proposed SC-GNN model with the baseline TISER-GCN and CNN models on the CI dataset when varying the input time windows from 5 seconds to 10 seconds. The results are presented in Table \ref{table:varying-time-windows}.

\begin{table}[h]
\centering
\begin{tabular}{|c|c|c|c|}
\hline
Time-Window & \textbf{Proposed SC-GNN} & TISER-GCN & CNN \\ \hline
10s         & 0.1137       & 0.2467 & 0.3593 \\ \hline
9s          & 0.1234       & 0.2619 & 0.3771 \\ \hline
8s          & 0.1406       & 0.2796 & 0.3988 \\ \hline
7s          & 0.1523       & 0.2968 & 0.4215 \\ \hline
6s          & 0.1608       & 0.3217 & 0.4503 \\ \hline
5s          & 0.1723       & 0.3604 & 0.4982 \\ \hline
\end{tabular}
\caption{MSE Comparison of the proposed SC-GNN model with baseline models on the CI dataset with varying input time windows.}
\label{table:varying-time-windows}
\end{table}

As the input time window is reduced, the performance of all models degrades, indicated by an increase in the MSE. However, the proposed SC-GNN model consistently outperforms the baseline TISER-GCN and CNN models across all input time windows. The SC-GNN model maintains a significantly lower MSE compared to the baselines, demonstrating its robustness and effectiveness in handling varying input sizes. Furthermore, the deterioration in the performance of the baseline models occurs much faster compared to our proposed SC-GNN when the input time window is shortened. Notably, even when using a 5s window input, the SC-GNN model demonstrates a remarkable 143\% improvement in performance compared to the next best-performing model, TISER-GCN, with a 10s input window. 

\subsubsection{Conditional Plots}
%Conditional Plots
\begin{figure*}[htbp]
    \centering
    % First row - Magnitude-based Conditional Scatter Plots
    \begin{subfigure}[b]{0.5\textwidth}
        \includegraphics[width=\linewidth]{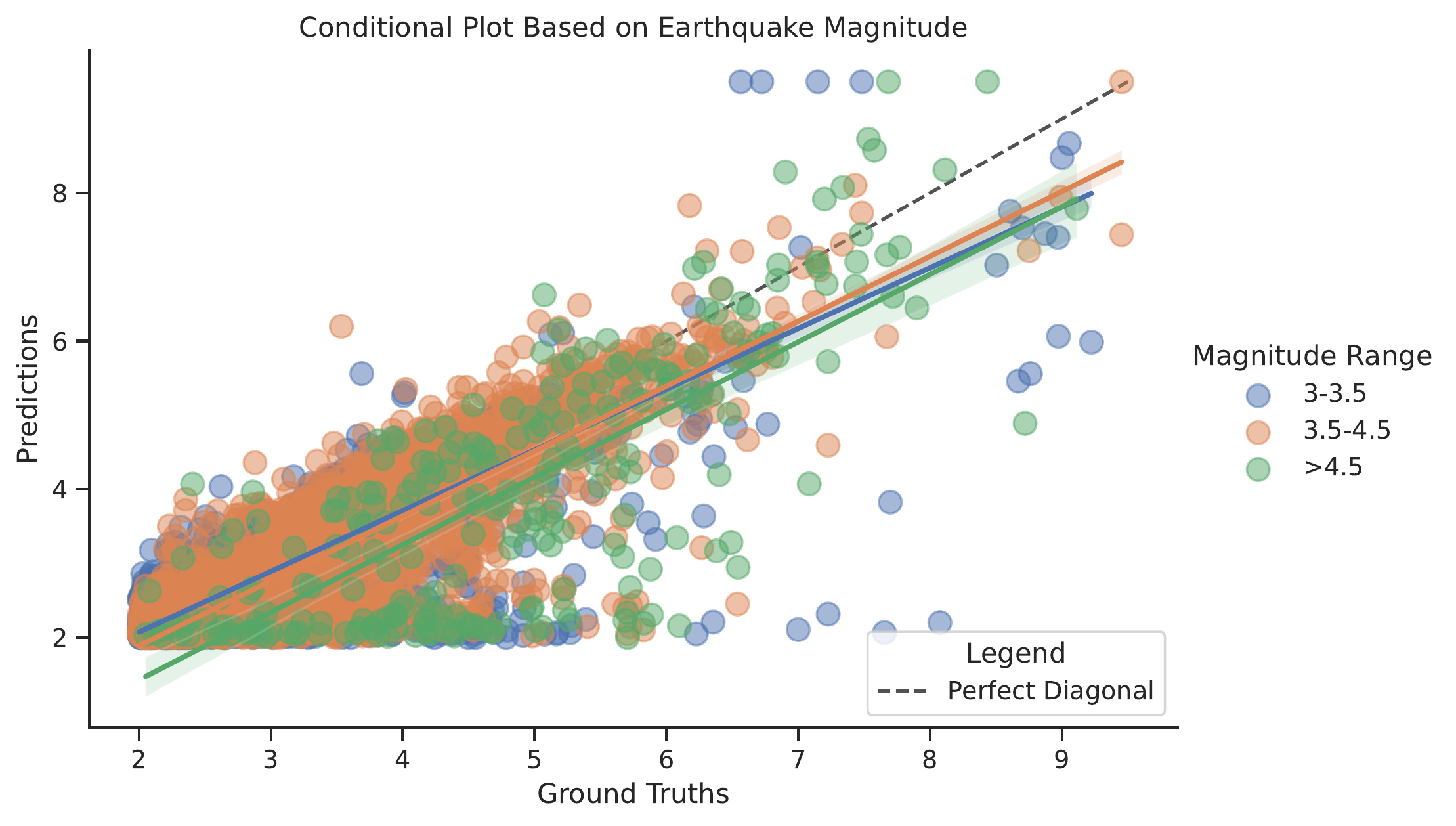}
        \caption{Proposed SC-GNN Model (Magnitude).}
        \label{fig: GNN Mag}
    \end{subfigure}\hfill
    \begin{subfigure}[b]{0.5\textwidth}
        \includegraphics[width=\linewidth]{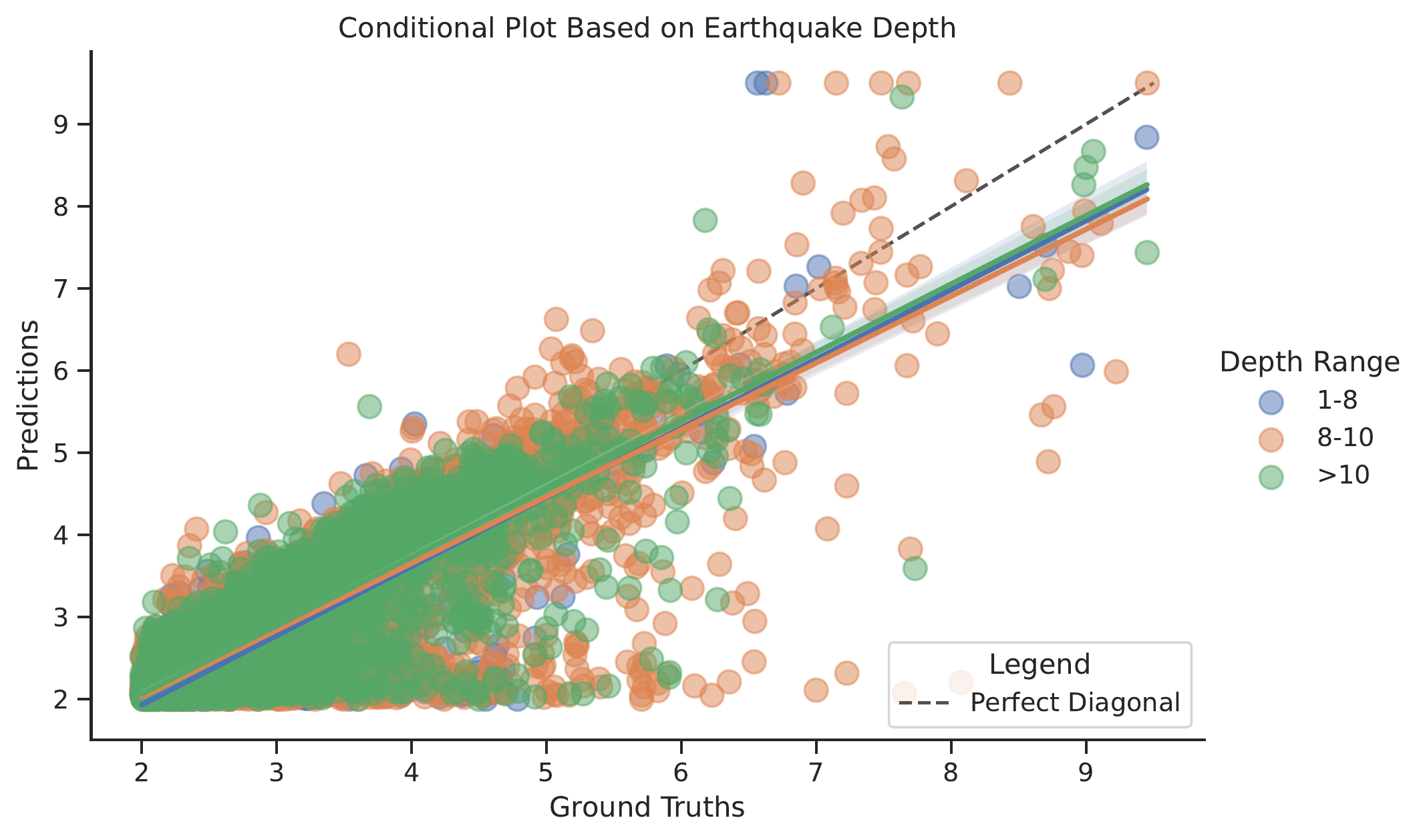}
        \caption{Proposed SC-GNN Model (Depth).}
        \label{fig: GNN Depth}
    \end{subfigure}
    
    % Second row - Magnitude-based Conditional Scatter Plots
    \begin{subfigure}[b]{0.5\textwidth}
        \includegraphics[width=\linewidth]{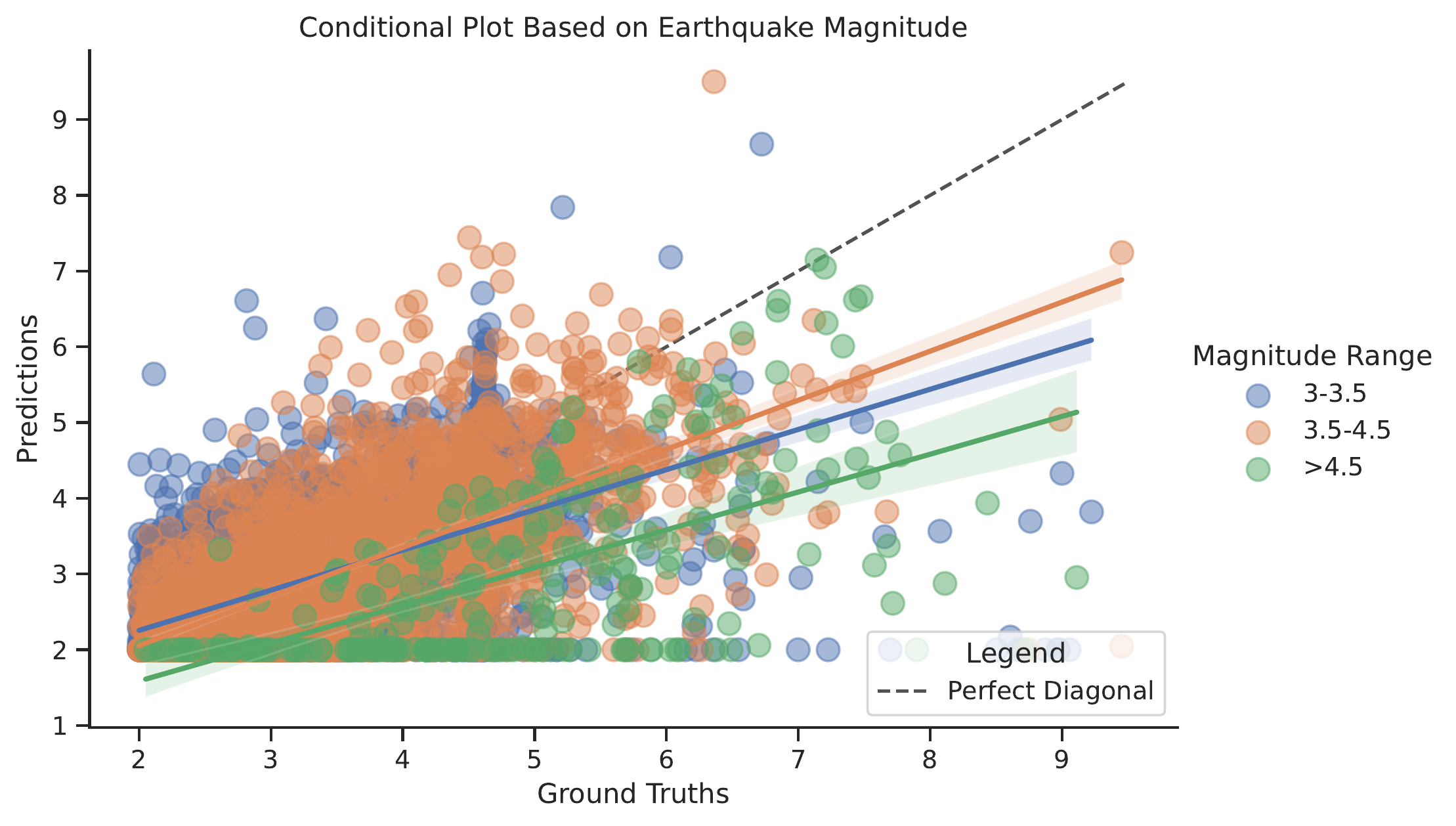}
        \caption{Baseline TISER-GCN Model (Magnitude).}
        \label{fig: GCN Mag}
    \end{subfigure}\hfill
    \begin{subfigure}[b]{0.5\textwidth}
        \includegraphics[width=\linewidth]{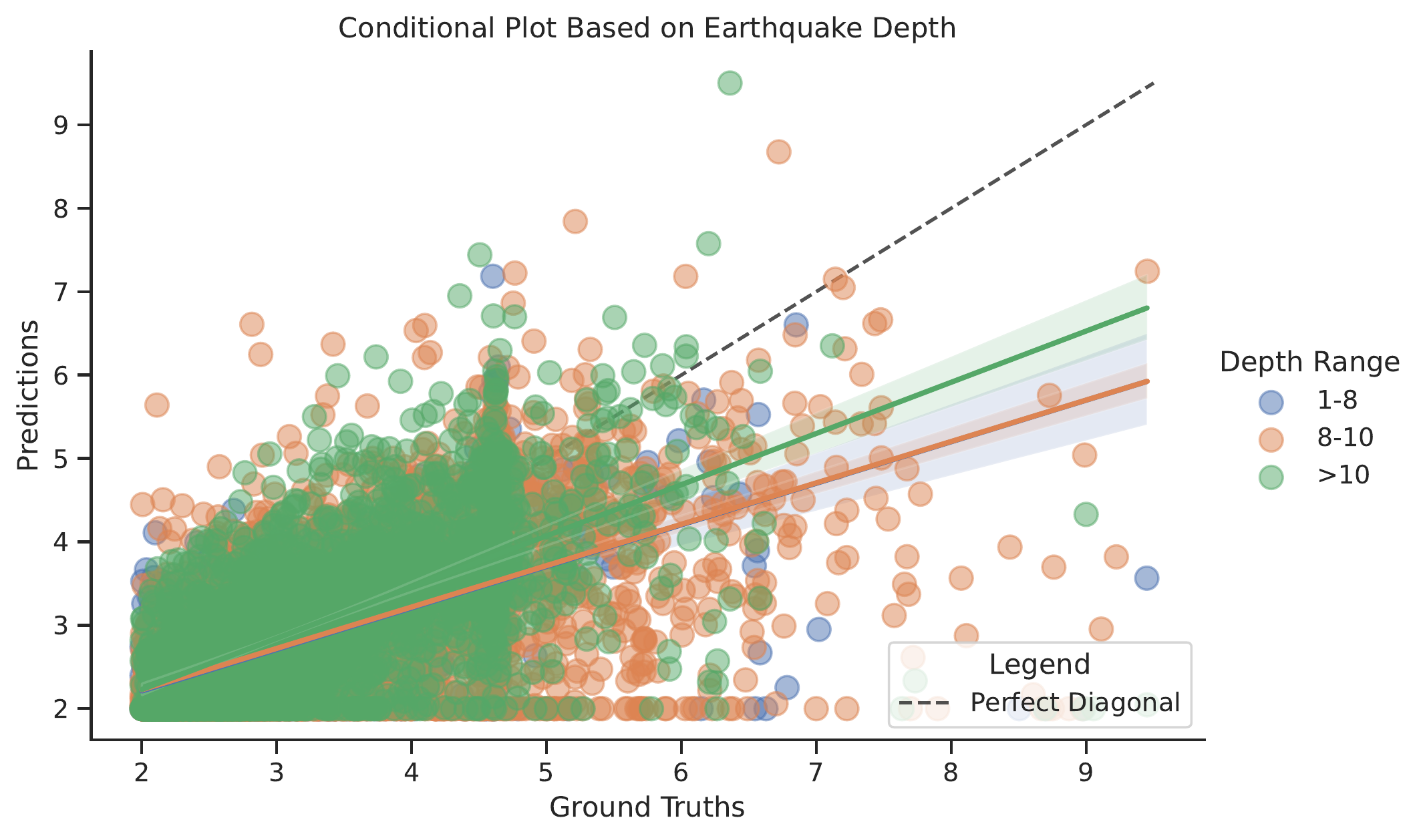}
        \caption{Baseline TISER-GCN Model (Depth).}
        \label{fig: GCN_depth}
    \end{subfigure}
    
    % Third row - Depth-based Conditional Scatter Plots
    \begin{subfigure}[b]{0.5\textwidth}
        \includegraphics[width=\linewidth]{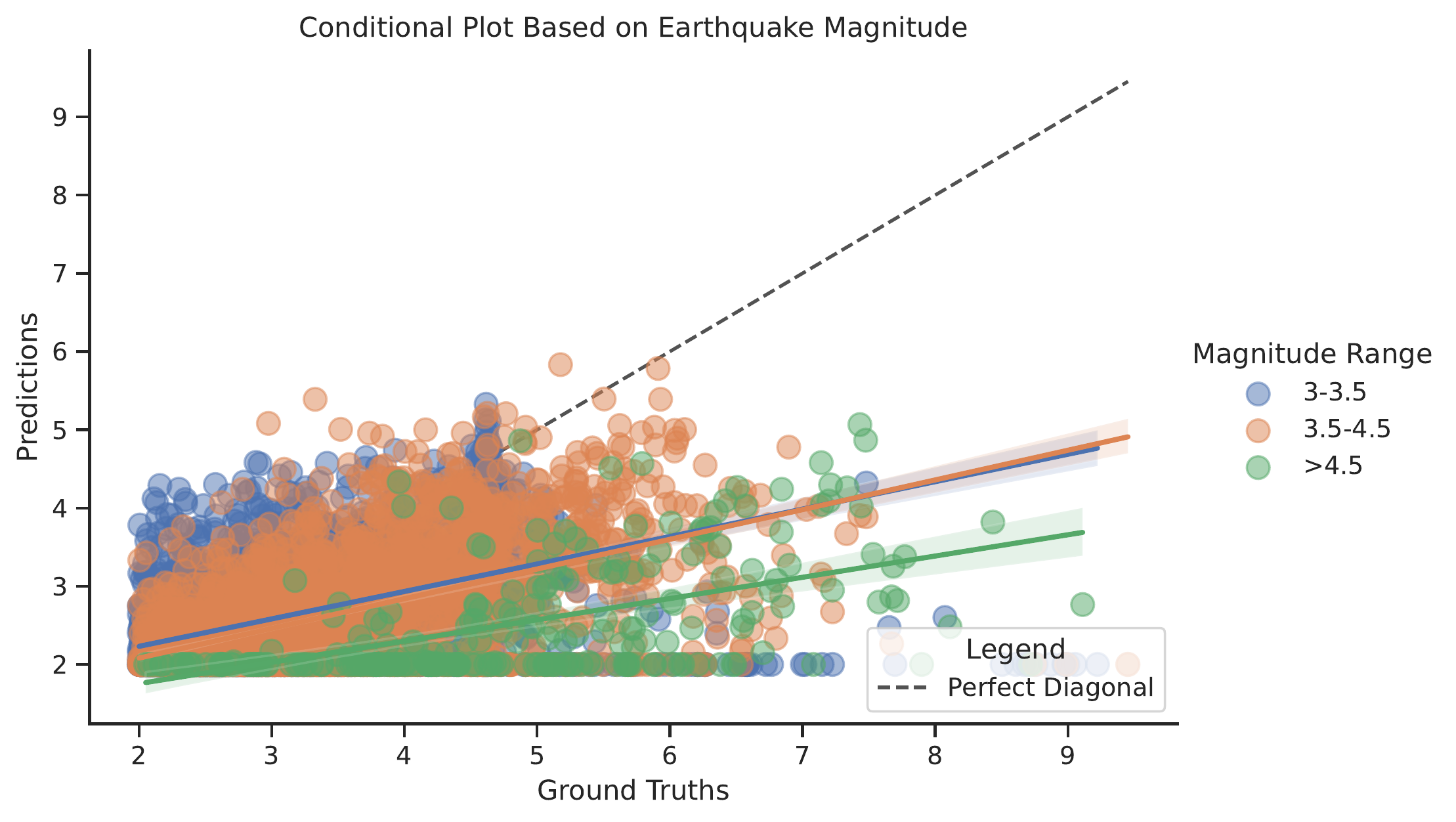}
        \caption{Baseline CNN Model (Magnitude).}
        \label{fig: CNN Mag}
    \end{subfigure}\hfill
    \begin{subfigure}[b]{0.5\textwidth}
        \includegraphics[width=\linewidth]{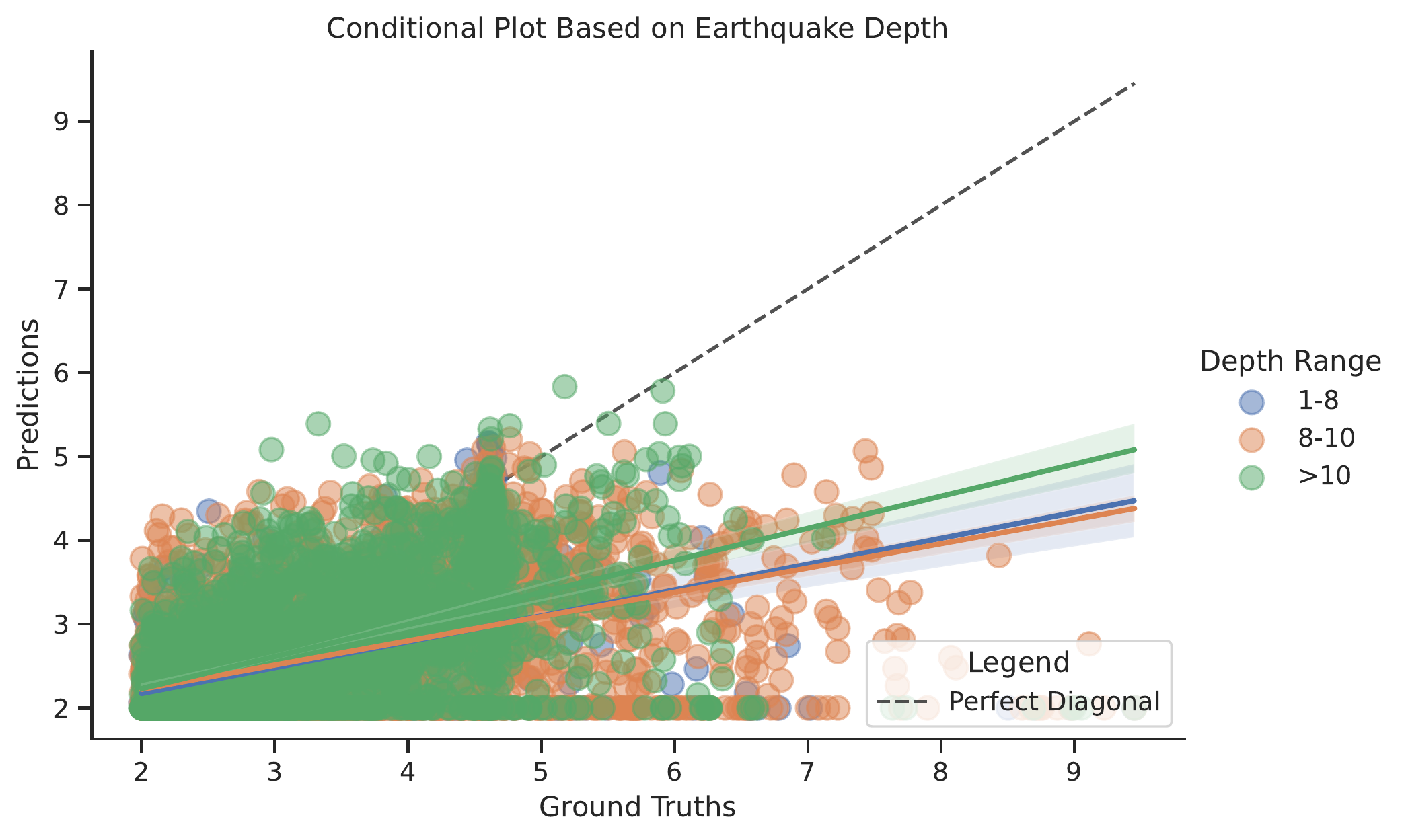}
        \caption{Baseline CNN Model (Depth).}
        \label{fig: CNN Depth}
    \end{subfigure}
    
    \caption{Conditional Scatter Plots: Magnitude-based (a,c,e) and Depth-based (b,d,f).}
    \label{fig: conditional_scatter_plots}
\end{figure*}

We have generated conditional scatter plots to better understand the performance of our SC-GNN model and the baseline models (TISER-GCN and CNN) concerning earthquakes' magnitude and depth. Conditional scatter plots based on the depth and magnitude of earthquakes help visualize the model's performance concerning earthquake depth and magnitude. By analyzing the plot, we can understand how well the model performs for different ranges of depth and magnitude, identify any potential biases, and assess the model's generalization capabilities.  Fig. \ref{fig: conditional_scatter_plots} (a,c,e) presents the magnitude-based conditional scatter plots for the proposed SC-GNN and baseline models. For the proposed SC-GNN model (Fig. \ref{fig: GNN Mag}), the regression lines for all magnitude ranges (3-3.5, 3.5-4.5, and greater than 4.5) almost overlap with each other, indicating that the predictions are unbiased with respect to the earthquakes' magnitudes. The regression lines are also quite close to the ideal regression diagonal line, suggesting good prediction accuracy across all magnitude ranges. It is imperative to stress that the intervals depicted in the conditional plot are derived from earthquake moment magnitudes, while the ground truth and our predictions are measured on an intensity scale.

In contrast, the baseline TISER-GCN model (Fig. \ref{fig: GCN Mag}) and CNN model (Figure \ref{fig: CNN Mag}) exhibit regression lines that do not overlap for all magnitude ranges, revealing biases in the predictions. Both models show much higher error for earthquakes with magnitudes greater than 4.5, with the regression lines being far from the ideal diagonal line. This indicates that the baseline models struggle to predict ground motion intensities for larger earthquakes accurately.

Fig. \ref{fig: conditional_scatter_plots}(b,d,f) illustrates the depth-based conditional scatter plots for the proposed SC-GNN and baseline models. For the proposed SC-GNN model (Fig. \ref{fig: GNN Depth}), the regression lines for all depth ranges (1-8 km, 8-10 km, and greater than 10 km) almost overlap with each other, signifying that the predictions are unbiased with respect to the earthquakes' depths. The regression lines are also quite close to the ideal regression diagonal line, demonstrating accurate predictions across various depths.

However, the baseline TISER-GCN model (Fig. \ref{fig: GCN_depth}) and CNN model (Fig. \ref{fig: CNN Depth}) exhibit regression lines that do not overlap for all depth ranges, highlighting biases in the predictions. Both models show much higher error for earthquakes with lower depths, with the regression lines being far from the ideal diagonal line. This implies that the baseline models have difficulty accurately predicting ground motion intensities for shallow earthquakes.

In summary, the proposed SC-GNN model outperforms the baseline models in terms of unbiased predictions and accuracy across different magnitude and depth ranges. These results further demonstrate the superiority of the SC-GNN model for ground motion intensity prediction.

\subsubsection{Bland-Altman Plots}

\begin{figure}[htbp]
    \centering
    \begin{subfigure}[t]{0.9\columnwidth}
        \includegraphics[width=\textwidth]{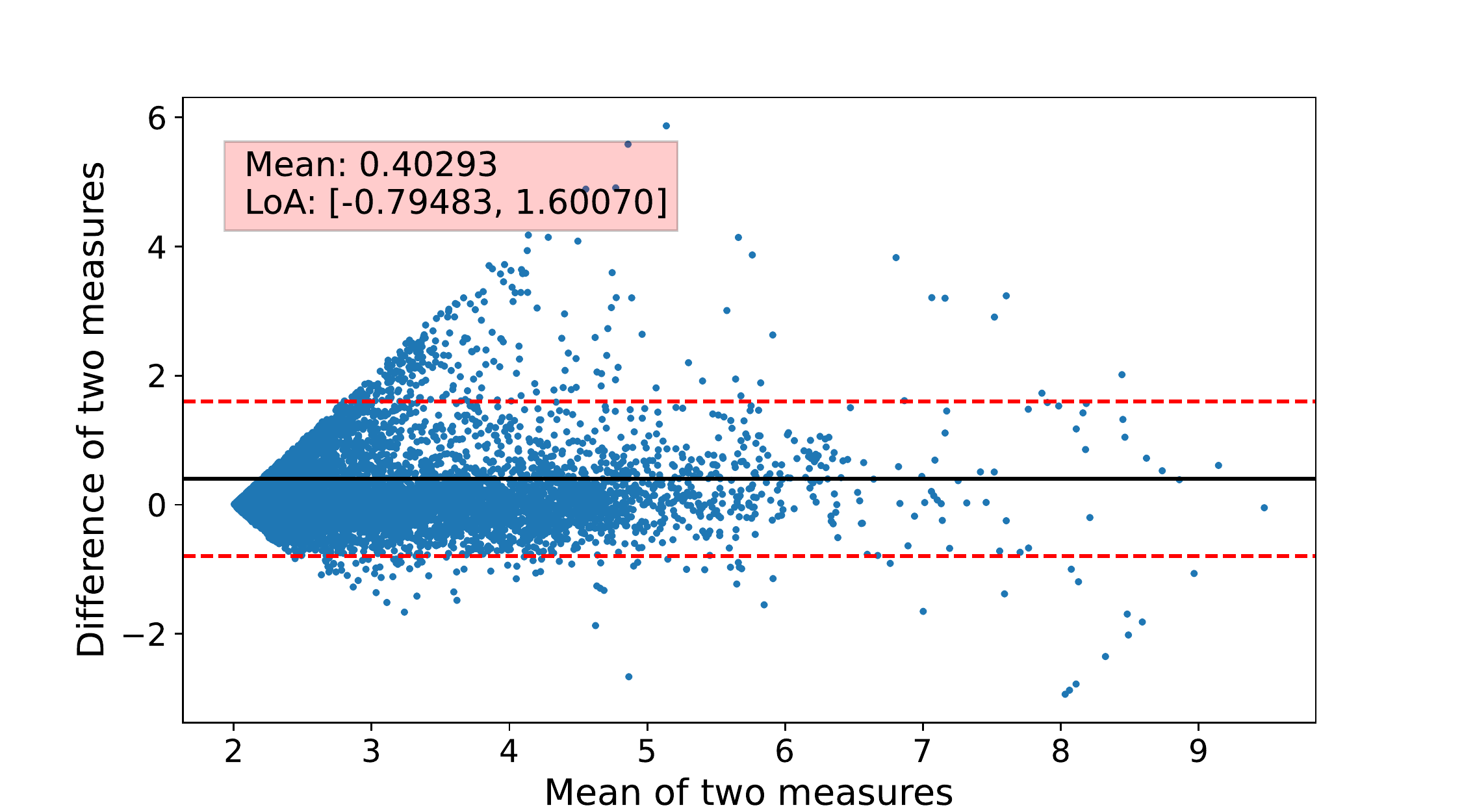}
        \caption{Proposed SC-GNN Model}
        \label{fig: GNN BA}
    \end{subfigure}
    \\
    \begin{subfigure}[t]{0.9\columnwidth}
        \includegraphics[width=\textwidth]{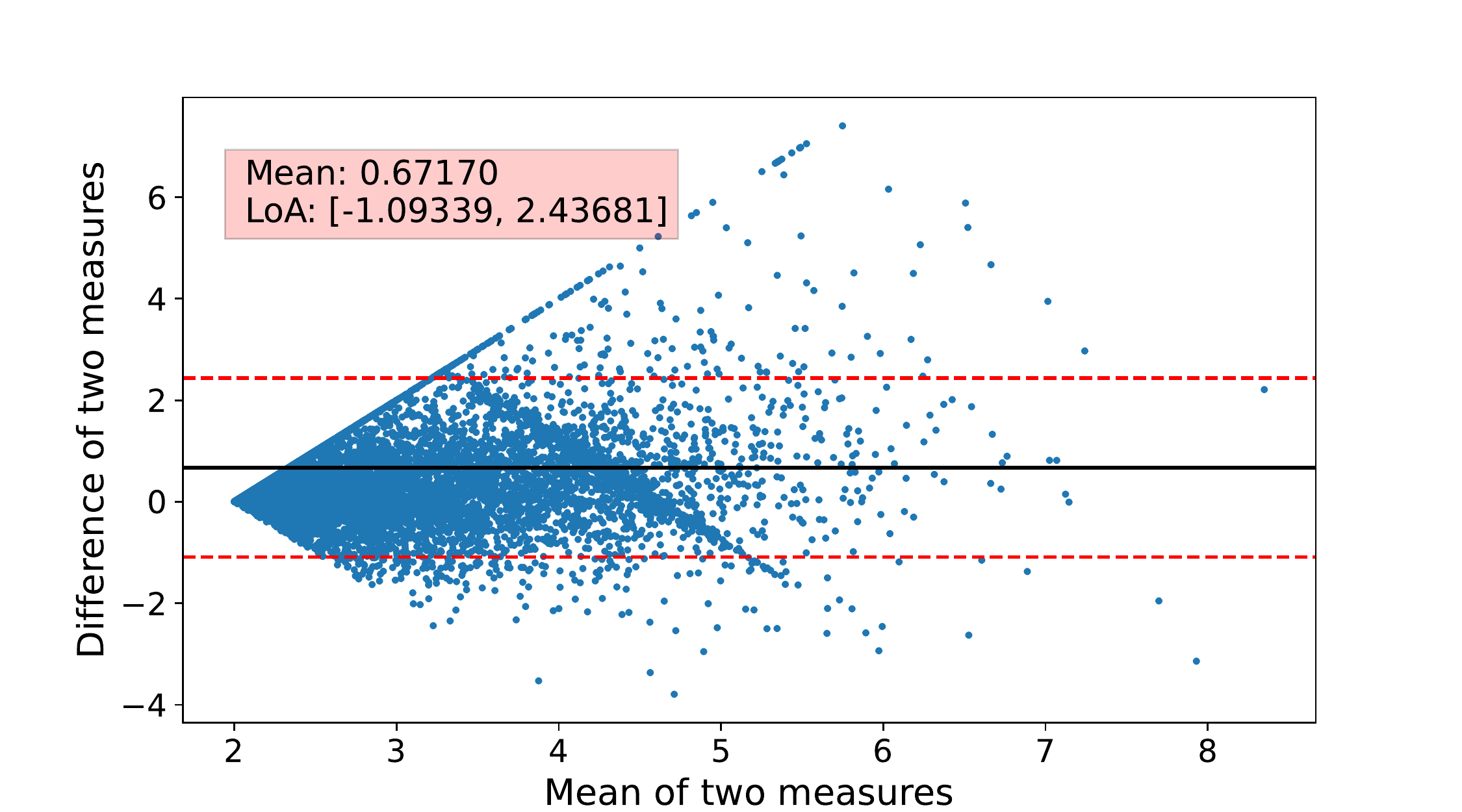}
        \caption{Baseline TISER-GCN Model}
        \label{fig: GCN BA}
    \end{subfigure}
    \\
    \begin{subfigure}[t]{0.9\columnwidth}
        \includegraphics[width=\textwidth]{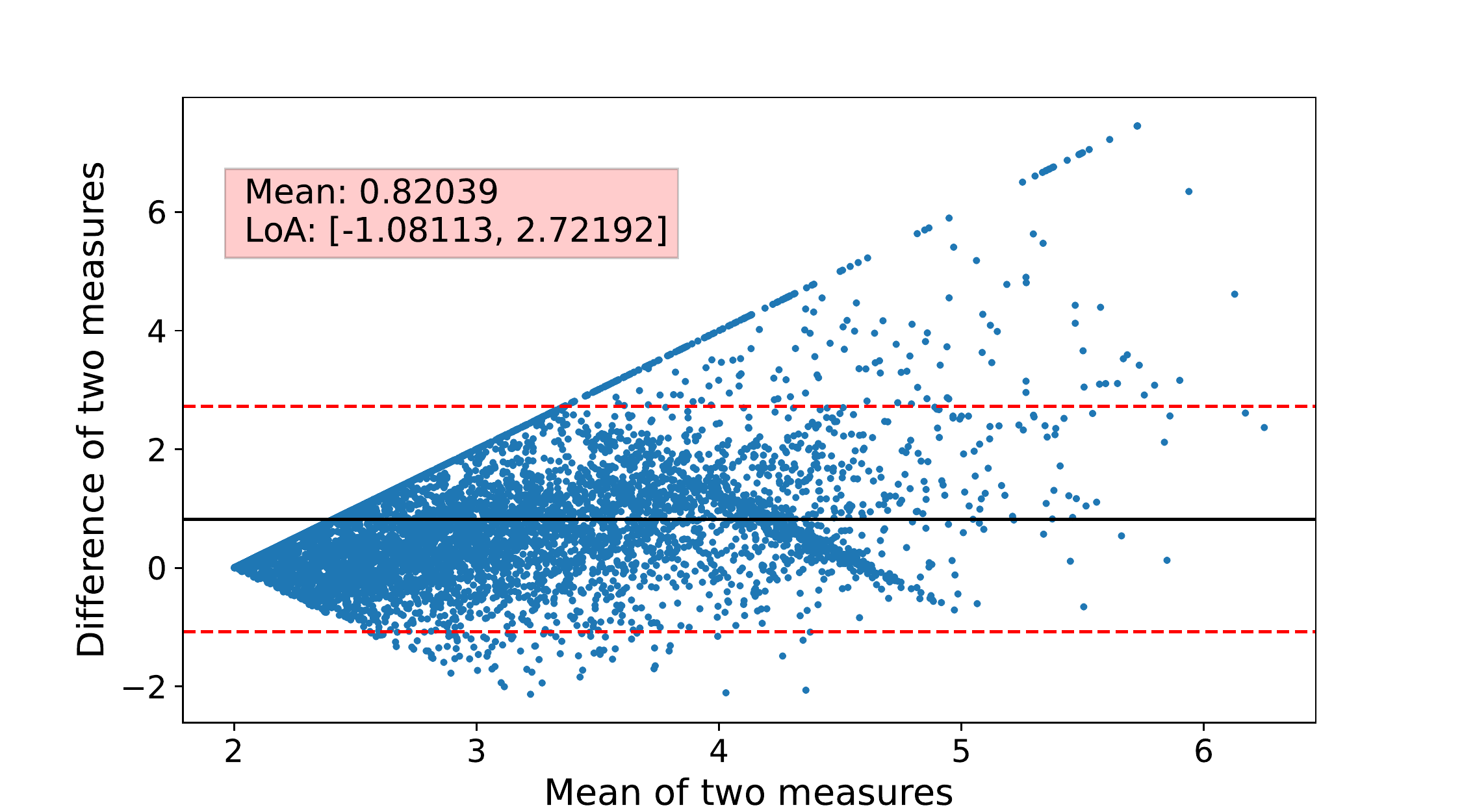}
        \caption{Baseline CNN Model}
        \label{fig: CNN BA}
    \end{subfigure}
    \caption{Bland-Altman plots displaying the mean difference and LOA of observed and predicted values.}
    \label{fig:bland_altman}
\end{figure}

 The Bland-Altman plots provide a useful visualization to assess the agreement between two different measurement techniques. In this case, we are comparing the predictions from the SC-GNN model and the baseline models (TISER-GCN and CNN) against the true observed earthquake intensities from the seismic waveforms. The main components of a Bland-Altman plot are the mean difference (bias) and the limits of agreement (LoA), which provide an estimate of the range within which 95\%  of the differences between the two measurements lie. The plots display the difference between the two methods against their average, allowing the identification of systematic biases, outliers, and trends in the differences. It aids in evaluating the model's consistency with respect to the actual observed intensity.

For the proposed SC-GNN model (Fig. \ref{fig: GNN BA}), the Bland-Altman plot shows a mean difference of 0.40 and LoA of [-0.79, 1.60]. This indicates that the SC-GNN model predictions are, on average, in good agreement with the true observed intensities. The narrow range of the LoA suggests that the model's performance is consistent across the range of earthquake intensities.

In contrast, the baseline TISER-GCN model (Fig. \ref{fig: GCN BA}) presents a mean difference of 0.67 and LoA of [-1.09, 2.43]. The increased mean difference compared to the SC-GNN model suggests that the TISER-GCN model predictions are less accurate. Additionally, the wider LoA indicates a higher level of variability in the model's performance.

For the baseline CNN model (Fig. \ref{fig: CNN BA}), the mean difference is 0.82, and the LoA are [-1.08, 2.72]. This result shows that the CNN model has the highest bias among the three models, with its predictions deviating significantly from the true observed intensities. The LoA is also wider than those for the SC-GNN and TISER-GCN models, suggesting a much greater level of variability in the performance of the CNN model.

In summary, the Bland-Altman plots demonstrate the superior performance of the proposed SC-GNN model in predicting earthquake intensities compared to the baseline TISER-GCN and CNN models. The GNN model exhibits the smallest mean difference and the narrowest LoA, indicating higher accuracy and consistency in its predictions across the range of earthquake intensities. This analysis further supports the effectiveness of the proposed SC-GNN model for predicting earthquake intensities.

\subsubsection{Performance Comparison on Secondary Datasets}
\begin{table}[h]
\centering
\begin{tabular}{|c|c|c|c|c|}
\hline
Metric/Model & \textbf{Proposed SC-GNN} & TISER-GCN & CNN & GMPE \\ \hline
MSE          & 0.5326       & 1.1664 & 1.7914 & 0.5765 \\ \hline
SD           & 0.6858       & 1.0163 & 0.9305 & 0.7539 \\ \hline
CC & 70.07\%  & 29.28\% & 16.94\% & 65.91\% \\ \hline
\( R^2 \) & 49.10\% & 8.57\% & 2.87\% & 43.44\% \\
\hline
\end{tabular}
\caption{Comparison of the proposed SC-GNN model with baseline models on the CW dataset for $2 < I_{\text{EMS-98}} < 9.5$.}
\label{table:cw-dataset}
\end{table}

\begin{table}[h]
\centering
\begin{tabular}{|c|c|c|c|c|}
\hline
Metric/Model & \textbf{Proposed SC-GNN} & TISER-GCN & CNN & GMPE \\
\hline
MSE & 0.8959 & 1.2196 & 1.5277 & 0.3512 \\
\hline
SD & 0.8295 & 0.9379 & 0.8682 & 0.5881 \\
\hline
CC & 40.68\% & 14.07\% & 13.66\% & 77.25\% \\
\hline
\( R^2 \) & 16.55\% & 1.98\% & 1.87\% & 59.68\% \\
\hline
\end{tabular}
\caption{Comparison of the proposed SC-GNN model with baseline models on the STEAD dataset for $2 < I_{\text{EMS-98}} < 9.5$.}
\label{table:stead-dataset}
\end{table}

In this discourse, we assess the performance of our proposed SC-GNN model on secondary datasets (CW and STEAD), comparing it to baseline models: TISER-GCN, CNN, and GMPE. Performance is gauged using MSE, SD, and CC.

In the evaluation on the CW dataset (Table \ref{table:cw-dataset}), the SC-GNN model demonstrates superior performance compared to the TISER-GCN and CNN models, as evidenced by lower MSE, SD, and higher CC. It also slightly outperforms the GMPE. It is worth noting that the CW dataset is smaller in size than CI in terms of the number of events and available waveform data, which leads to a decrease in the performance of all DL-based models.

In the case of the STEAD dataset (Table \ref{table:stead-dataset}), the SC-GNN model exhibits significant improvements over all the baseline models except for the GMPE. This is reflected in lower MSE, lower SD, and higher CC values. However, it is important to mention that the STEAD dataset comprises only 4299 waveform data points, accounting for a mere 11.60\% of the possible 37054 waveforms ($191\times 194$). Consequently, the limited data availability poses challenges for these data-driven DL models, including the proposed SC-GNN, whereas the GMPE remains relatively unaffected. Nonetheless, the SC-GNN model showcases comparatively better resilience and predictive capabilities on this challenging STEAD dataset, further affirming its effectiveness in earthquake ground shaking intensity prediction.

\subsubsection{Comparison of Model Parameters}
We carry out a detailed analysis of the number of parameters utilized by our proposed SC-GNN and compare this to the parameter count of two other baseline models: the TISER-GCN and the CNN. A summary of these comparisons is presented in Table \ref{table:comparison_of_models}.

\begin{table}[h]
\centering
\begin{tabular}{|c|c|c|}
\hline
Model & Parameters (Millions) & Model Size (MB) \\
\hline
\textbf{Proposed SC-GNN} & 0.705 & 2.8 \\
\hline
TISER-GCN & 1.26 & 4.8 \\
\hline
CNN & 1.35 & 5.1 \\
\hline
\end{tabular}
\caption{Comparison of model parameters and model sizes.}
\label{table:comparison_of_models}
\end{table}

As we observe, our proposed SC-GNN model utilizes fewer parameters, approximately 0.705 million, and consequently, its overall model size is 2.8 MB. In contrast, both the TISER-GCN and CNN models require almost twice as many parameters, around 1.26 and 1.35 million, respectively, and larger model sizes of 4.8 and 5.1 MB.

The significance of these findings becomes even more pronounced in the context of early warning systems for seismic activities. A lower number of parameters directly implies a more efficient model in terms of computational resources and thus faster computations, which is a critical factor in timely predicting seismic activities and issuing early warnings.

Furthermore, the reduced model size of the SC-GNN makes it a more suitable choice for implementation on resource-constrained devices. This characteristic is critical, as seismic early warning systems often operate on field-deployed devices with limited computational power and storage capacity.

These factors underscore the suitability of SC-GNN over the TISER-GCN and CNN models in the context of seismic activity prediction. They also validate the design choices in developing the SC-GNN, emphasizing its efficient utilization of parameters without sacrificing prediction performance, as evidenced in the earlier discussions of model accuracy and performance on varying input window lengths.

\subsection{Ablation Study}
 \subsubsection{Model Variations}

 In this section, we present the results of an ablation study conducted to evaluate the impact of different combinations of layers in our proposed SC-GNN model. The final proposed model consists of a combination of two Chebyshev Conv (ChC), one Graph-skip Conv (GCSC), and one Graph-Attention Pool (GAP) layer. We experimented with removing some layers and adding extra layers, such as the graph convolutional layer (GCN), Diffusion Conv (DC), and graph attention layer (GAT), to demonstrate that the final proposed model performs better than other layer combinations. The results are shown in Table \ref{table:ablation-study}.

\begin{table}[h]
\centering
\begin{tabular}{|l|c|c|}
\hline
GNN Layers & MSE & Normalized MSE \\ \hline
\textbf{ChC + ChC + GCSC + GAP} & 0.1137 & 4.83\% \\ \hline
ChC + ChC + GCSC & 0.1848 & 7.86\% \\ \hline
ChC + ChC + GAT & 0.5431 & 23.11\% \\ \hline
ChC + ChC + GCN & 0.4337 & 18.45\% \\ \hline
ChC + ChC + DC & 0.9372 & 39.88\% \\ \hline
ChC + GCSC + GAP & 0.3697 & 15.73\% \\ \hline
\end{tabular}
\caption{Ablation study results for the proposed SC-GNN model on the CI dataset.}
\label{table:ablation-study}
\end{table}

The results from the ablation study clearly demonstrate that the final proposed model, which combines ChC, GCSC, and GAP layers, achieves the lowest MSE (0.1137) and the lowest normalized MSE (4.83\%). This indicates that the combination of these layers is the most effective in predicting ground motion intensities. Here, the normalized MSE is obtained by dividing the raw MSE values by the observed data mean. This computational step guarantees that the normalized MSE appropriately reflects the relative error, taking into consideration the data set scale.

The superior performance of the final proposed model can be attributed to the combined strengths of the ChC, GCSC, and GAP layers. The ChC layer effectively captures local spatial information in the graph, while the GCSC layer helps learn long-range dependencies and skip uninformative features. The GAP layer, on the other hand, focuses on aggregating the most relevant information from the graph by attending to the most important nodes.

By comparing the final proposed model with other layer combinations, we can infer that removing any of these layers leads to decreased performance, as evidenced by higher MSE and normalized MSE values. This confirms that the synergy between ChC, GCSC, and GAP layers is crucial for achieving the best performance in our GNN model.
 
\subsubsection{Time Window Variation}
We evaluated the performance of the proposed SC-GNN model by varying the input window length from 5s to 30s. The results of this analysis are presented in Table \ref{table:window-length-performance}.

\begin{table}[h]
\centering
\begin{tabular}{|c|c|c|c|}
\hline
Time-Window & MSE & SD & CC \\ \hline
30s & 0.0844 & 0.2893 & 92.86\% \\ \hline
25s & 0.0880 & 0.2985 & 92.46\% \\ \hline
20s & 0.0969 & 0.3107 & 91.89\% \\ \hline
15s & 0.1030 & 0.3181 & 91.38\% \\ \hline
10s & 0.1137 & 0.3369 & 90.09\% \\ \hline
5s & 0.1723 & 0.4071 & 83.29\% \\ \hline
\end{tabular}
\caption{Performance analysis of the proposed SC-GNN model with varying input window length}
\label{table:window-length-performance}
\end{table}

\begin{figure*}[! h]
\centering
    \includegraphics[width=2.1\columnwidth,height=45mm]{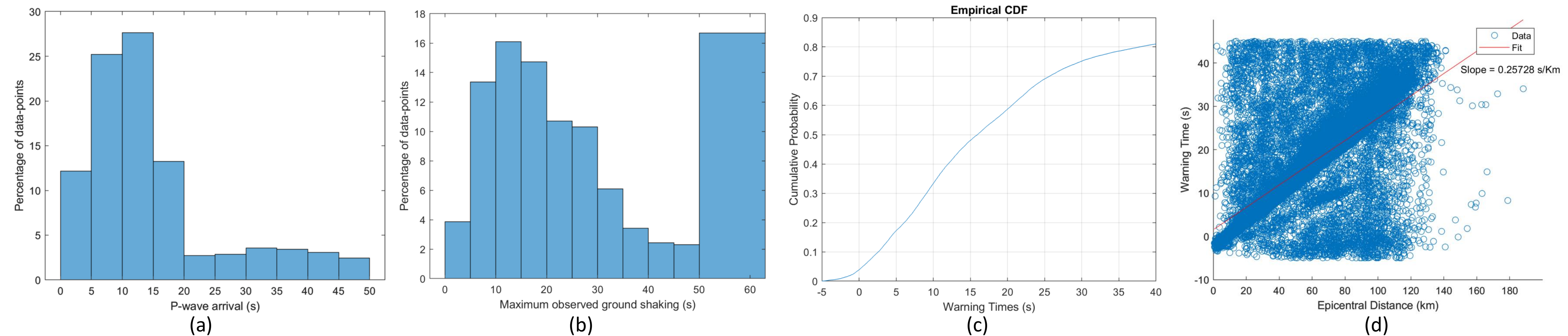}
    \caption{SC-GNN for early warning: a) Histogram of p-wave arrival times. b) Histogram of the instant at which maximum ground shaking occurs. c) Cumulative density function of warning times. d) Plot of warning time with distance from the epicenter.}
    \vspace{-2mm}
    \label{Fig_EEW}
\end{figure*}

As shown in the table, the performance of the SC-GNN model deteriorates as the input window length is reduced, with the MSE increasing and the CC decreasing. The reason for this deterioration is that shorter windows capture less information about the seismic waveforms, making it more challenging for the model to accurately predict ground motion intensities.

Earthquake early warning systems (EEWS) require shorter window lengths for faster response times, which is crucial for timely alerts and potentially saving lives and property. However, as demonstrated in our analysis, there is a trade-off between window length and prediction accuracy. The shorter the window length, the less accurate the model becomes.

To achieve an optimal balance between response time and prediction accuracy for EEWS applications, it is essential to carefully consider the choice of input window length. Future research could focus on further optimizing the GNN model or exploring other approaches that maintain high accuracy while using shorter input windows.

\subsection{SC-GNN for Early Warning}

In this sub-section, we showcase the significant promise and effectiveness of our proposed model, SC-GNN, as an integral part of the earthquake early warning (EEW) systems. The ability of the SC-GNN to deliver rapid and accurate seismic intensity predictions, even in the critical window of just 5s, underscores its potential as a pioneering tool for early seismic warnings. The following discussion contains some key results using the main CI dataset.

A vital testament to the utility of SC-GNN in EEW is reflected in the distribution of the P-wave arrival times depicted in Fig. \ref{Fig_EEW}(a). The histogram reveals that a majority of the stations, approximately 90\%, receive the prediction before the P-wave arrives \footnote{Here, we assume 5s window input to the SC-GNN.}, considering that our SC-GNN, a regional EEW system, enables early warning prior to the arrival of P-waves at stations located at greater distances. It is important to note that the actual warning time available to take preparatory measures will also depend on the time required to disseminate warnings, which can vary based on infrastructure, technology, and geolocation. Research shows that, with ideal infrastructure, this transmission time can be relatively negligible compared to the early detection advantage provided by SC-GNN \cite{pieska2017tcp,jiang2018low}.

The histogram of the maximum ground shaking times, as shown in Fig. \ref{Fig_EEW}(b), further underscores the benefit of SC-GNN. As observed, more than 95\% of the locations potentially receive the warning well ahead of the maximum ground shaking, often the most destructive phase of an earthquake. This suggests that taking into account practical transmission times, there might still be a valuable window for the populace and infrastructure to prepare, potentially mitigating the seismic event's impacts.

The cumulative density function (CDF) of the warning times (Fig. \ref{Fig_EEW}(c)) provides an illustrative perspective on the capabilities of SC-GNN in EEW. The plot suggests that timely warnings could be disseminated to a significant number of areas, strengthening the argument for integrating SC-GNN into EEW systems. We observe that around 70\% of the locations potentially receive a warning time of more than 10s, which, after accounting for transmission times, might be sufficient for various precautionary measures like taking cover, shutting off utilities, evacuation, etc \cite{minson2018limits}. Here, \emph{warning time} refers to the interval between the occurrence of maximum ground shaking and the moment the warning is received.

Furthermore, the relationship between the warning time and the epicentral distance (Fig. \ref{Fig_EEW}(d)) affirms the effectiveness of SC-GNN. The warning time proportionally increases with the distance from the epicenter; for approximately every 4 km, the warning time is incremented by 1s. This suggests that areas farther from the epicenter, which traditionally had to wait longer for the warning will now have more time to brace for the incoming seismic waves.

In summary, while acknowledging the practical considerations of warning transmission times, our proposed SC-GNN framework exhibits promising potential for integration into EEW systems. By leveraging the initial seismic waveforms, SC-GNN aims to extract critical earthquake information and generate accurate seismic intensity predictions, contributing to early warning efforts. It is anticipated that this could significantly increase the warning lead times in a majority of areas, providing a valuable cushion of time for implementing appropriate disaster mitigation measures.

\section{Conclusion}
In this paper, we have proposed a novel deep learning framework, SC-GNN, that comprises two key deep learning components: a GNN for capturing spatiotemporal characteristics of seismic waves in a geographical area, and a contrastive learning module to find the representation of seismic waves from a small portion of initial seismic waveforms. More specifically, the GNN part has a unique ability to propagate information through the nodes of a graph-like structure of seismic station distribution, where wave propagation enables globally informed predictions with locally available data.
On the other hand, the self-supervised contrastive learning phase enabled us to learn the representation of seismic waveforms in such a way that facilitates predicting seismic intensity from a significantly shorter input waveform, which is a key factor in an EEW system. We have shown in experiments that the proposed SC-GNN is adaptive to varying input window lengths, with a commendable performance even at a reduced window length of 5s. This trait is particularly valuable for EEW systems, where every second of early warning can mean the difference between life and death. Our SC-GNN model outperformed all state-of-the-art methods on three well-known seismic datasets across multiple assessment measures. Finally, when potentially integrated into an EEW system, SC-GNN delivers rapid and accurate seismic intensity predictions, with approximately 90\% of the stations receiving warnings even before the P-wave arrival.

Incorporating meta-learning and transfer-learning schemes and exploring their potential applications within the broader field of seismic event prediction and disaster management can be a potential future research.

%\textbf{Data and Code Availability:} Data and code are available here.

\textbf{Data and Code Availability:} For reproducibility of the results, data and code are made available at Github: \url{https://github.com/Blaze-raf97/Seismic-Intensity-Prediction-using-SC-GNN-for-EEW}.

%\appendices
%\%section{Proof of the First Zonklar Equation}

%\section{}

\section*{Acknowledgment} This project is funded by
RISE Internal Research Project ID 2021-01-027, Title: ``Earthquake Early Warning System in Bangladesh", from Bangladesh University of Engineering and Technology (BUET), Dhaka.


\ifCLASSOPTIONcaptionsoff
  \newpage
\fi



%\bibliographystyle{IEEEtran}


% \end{thebibliography}

% \normalem
\bibliographystyle{IEEEtran}
\bibliography{Main_file}


%\newpage

\begin{IEEEbiography}[{\includegraphics[width=1in,height=1.25in,clip,keepaspectratio]{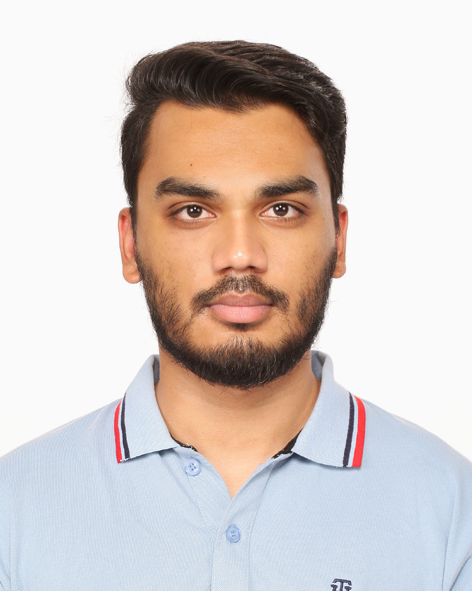}}]{Rafid Umayer Murshed} is a researcher and engineer with a background in Electrical and Electronics Engineering (EEE). He obtained his BSc degree in EEE from Bangladesh University of Engineering and Technology (BUET) in May 2022. Currently, he is pursuing his master's degree in the Electrical and Computer Engineering (ECE) department at the University of Texas at Dallas under the supervision of Dr. Mohammad Saquib. He focuses on developing innovative deep learning-based approaches for 6G and Beyond wireless communication networks. His research interests include UM-MIMO, beamforming, resource allocation, reconfigurable intelligent surfaces, remote sensing, signal processing, and wireless health monitoring. His scholarly work has been published in renowned journals and conference proceedings. %, reflecting the quality and impact of his research endeavours.

%Mr Murshed has made noteworthy contributions to the field of wireless communication technologies, particularly in the application of deep learning and reinforcement learning. His research interests include UM-MIMO, beamforming, resource allocation, reconfigurable intelligent surfaces, remote sensing, signal processing, and the Internet of Things (IoT). His scholarly work has been published in renowned journals and conference proceedings, reflecting the quality and impact of his research endeavours.
\end{IEEEbiography}

\begin{IEEEbiography}
[{\includegraphics[width=1in,height=1.25in,clip,keepaspectratio]{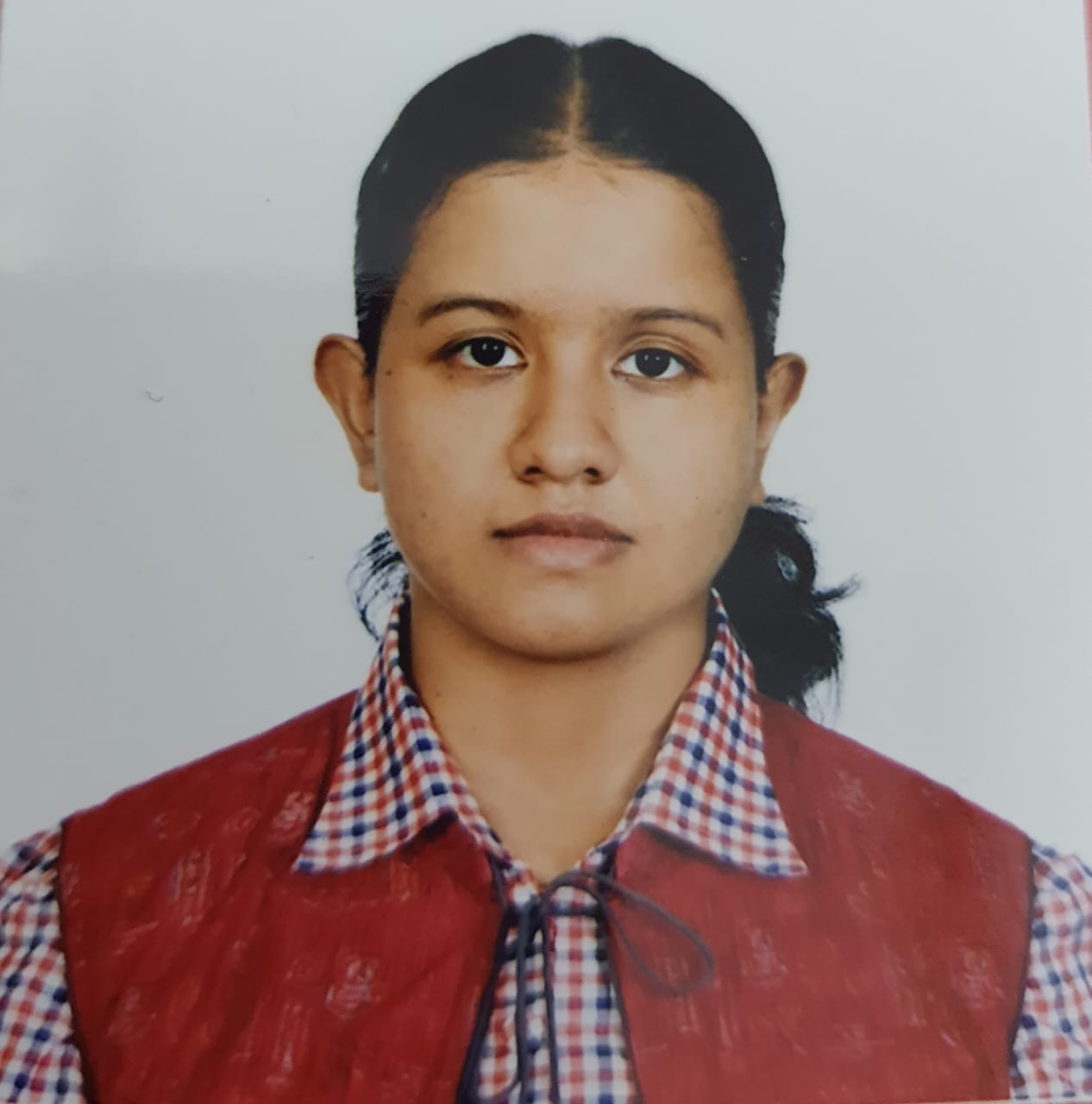}}]{Kazi Noshin} completed her B.Sc. from the Department of Computer Science and Engineering at Bangladesh University of Engineering and Technology (BUET) in May 2022. She holds the position of a research assistant at the BUET-Japan Institute of Disaster Prevention and Urban Safety (JIDPUS). Her primary responsibilities at JIDPUS include preparing seismic datasets and developing deep learning-based models to improve earthquake early warning systems. Her research interest primarily focuses on Artificial Intelligence, Machine Learning, and Deep Learning techniques, concerned with human life improvements, health, and behavior. %By harnessing the power of AI and machine learning, she aspires to develop innovative solutions that can positively impact society, healthcare, and behavior analysis.
\end{IEEEbiography}

\begin{IEEEbiography}
[{\includegraphics[width=1in,height=1.25in,clip,keepaspectratio]{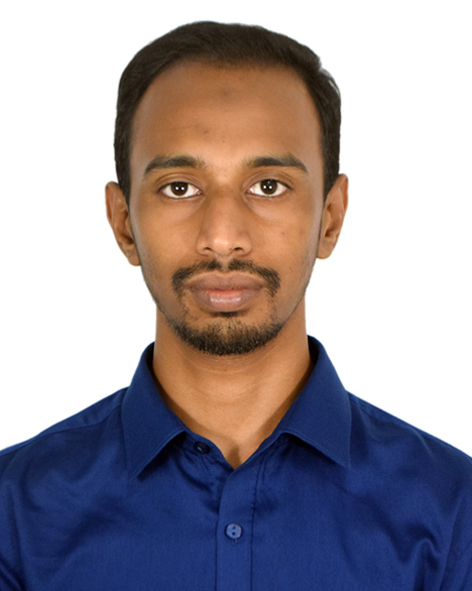}}]{Md. Anu Zakaria} holds a B.Sc. degree in Civil Engineering from Bangladesh University of Engineering and Technology (BUET), obtained in May 2022. He is currently pursuing his M.Sc. degree in Structural Engineering at BUET, with a particular focus on seismic assessment of structures. As a Research Assistant at BUET-JIDPUS, he is involved in the project  ``Earthquake Early Warning System in Bangladesh", where his primary responsibility is to prepare seismic and geotechnical data to be used for machine learning techniques.  His research interests revolve around dynamic analysis of structures, concrete technology and finite element modeling of concrete bridges.

%Mr. Zakaria is experienced in the modeling and seismic analysis of prestressed concrete girder bridges. His research interests revolve around dynamic analysis of structures, concrete technology, and bridge design. With his knowledge in these areas, he looks forward to contribute in the advancement of the field of Structural and Earthquake Engineering.
\end{IEEEbiography}

% Others bios need to be added here.

\begin{IEEEbiography}[{\includegraphics[width=1in,height=1.25in,clip,keepaspectratio]{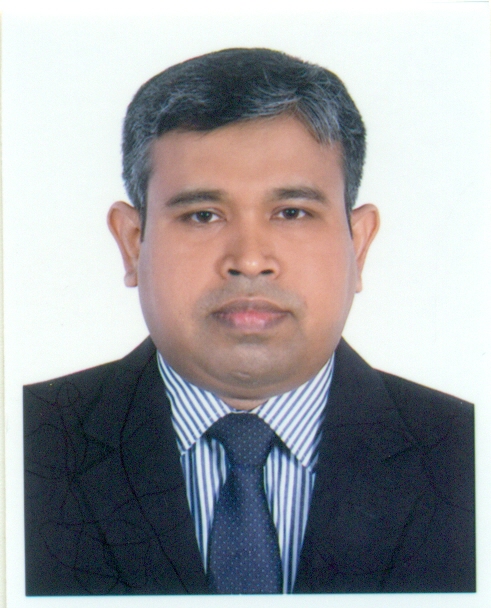}}]{Md. Forkan Uddin (M’13)} received his B.Sc. and M.Sc. degrees both in Electrical and Electronic Engineering (EEE) from Bangladesh University of Engineering and Technology (BUET), Dhaka, Bangladesh in 2003 and 2005, respectively, and his Ph.D. degree in Electrical and Computer Engineering from University of Waterloo, Canada in 2011. In 2003, he joined as a Lecturer in the Department of EEE, BUET, Bangladesh, where he is currently a Professor. His research interests include wireless communications and networking, smart grid communications and load management, and optical fiber communications. Dr. Uddin is a member of the Editorial Board of the Journal of Institutional Engineering and Technology. He received ``UGC Gold Medal 2017" in the category ``Engineering and Technology" from the University Grants Commission (UGC) of Bangladesh for his outstanding contributions to fundamental research work.% He is a co-recipient of the Dr. Faterma Rashid Best Paper Awards in the 8th and 9th International Conference on Electrical and Computer Engineering (ICECE) in 2014 and 2016. He also received the Best Paper Award Communication in the first IEEE International Conference on Telecommunications and Photonics (ICTP) 2015. He has been worked in many national projects related to power and communication systems in Bangladesh. He has been served in several positions of Executive Committees of IEEE Bangladesh Section and its Communications Society Chapter.
\end{IEEEbiography}

\begin{IEEEbiography}[{\includegraphics[width=1in,height=1.25in,clip,keepaspectratio]{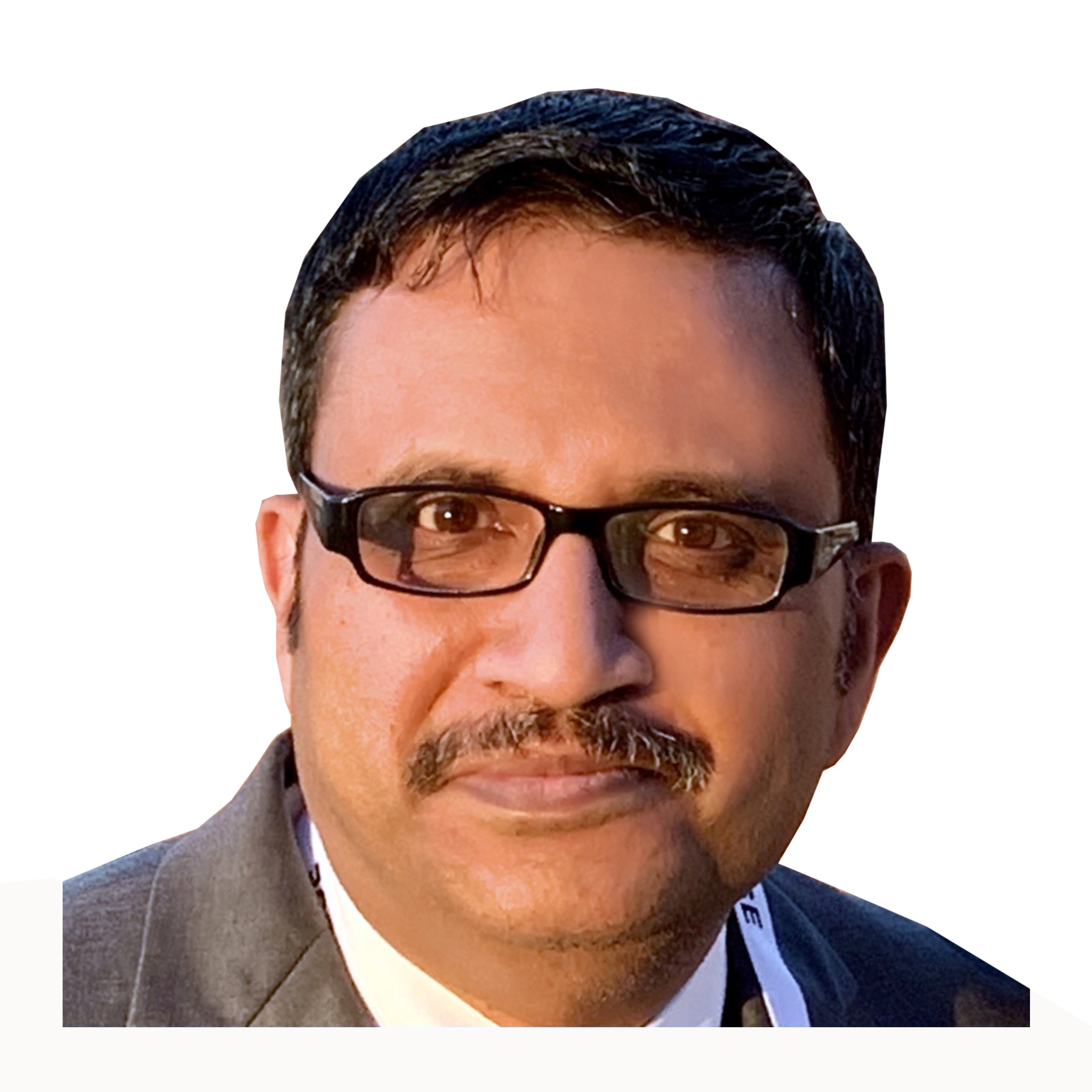}}]{A. F. M. S. Amin} is a Professor in the Department of Civil Engineering and the Director at BUET-Japan Institute of Disaster Prevention and Urban Safety of Bangladesh University of Engineering and Technology. His academic and research interests include teaching in civil engineering; he has led dedicated research and development teams to design and oversee the repair and retrofitting of structures, earthquake protection, helping increase and preserve the structural soundness of many important buildings and remarkable bridges in Bangladesh.
\end{IEEEbiography}

\begin{IEEEbiography}[{\includegraphics[width=1in,height=1.25in,clip,keepaspectratio]{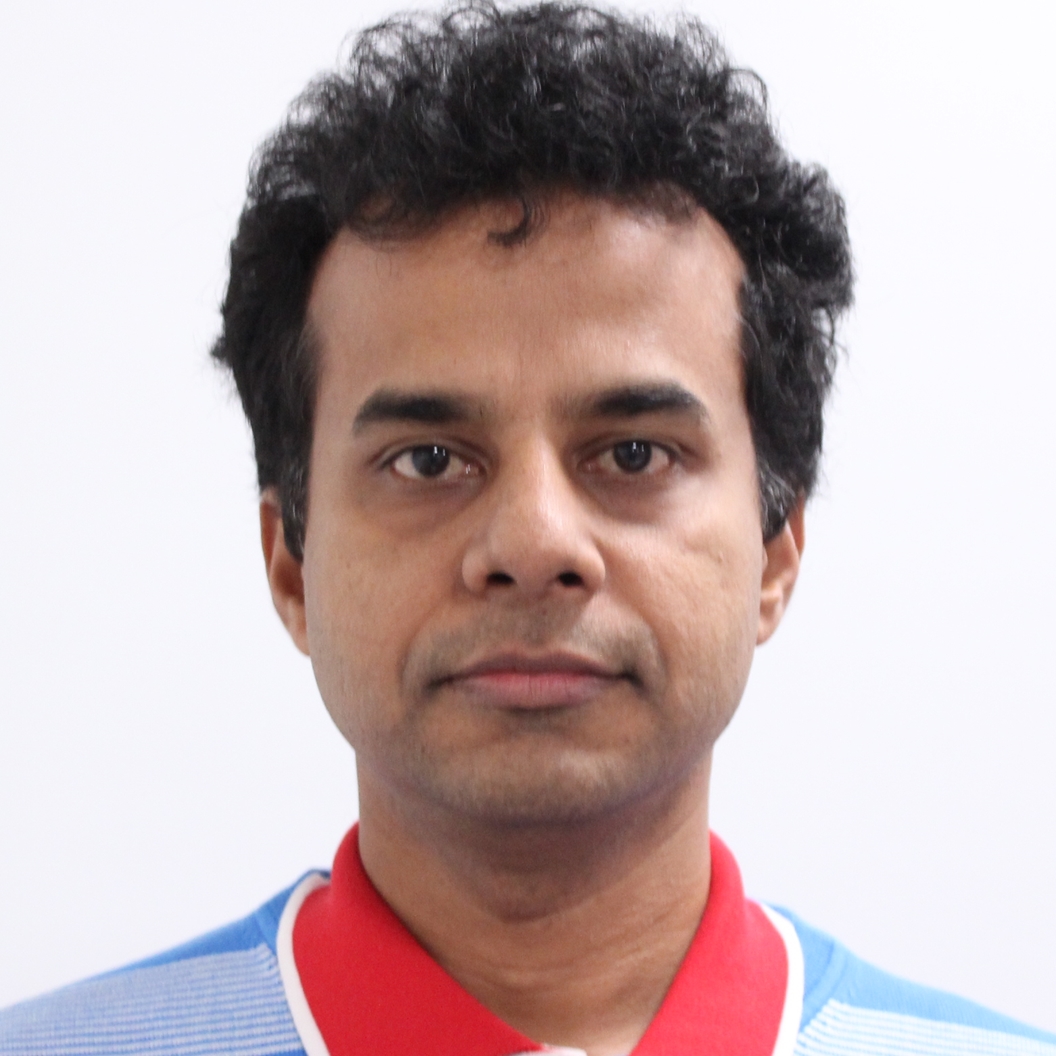}}]{Mohammed Eunus Ali} is a Professor in the Department of Computer Science and Engineering (CSE) at Bangladesh University of Engineering and Technology (BUET). He is the group leader of Data Science and Engineering Research Lab (DataLab) at CSE, BUET. His research areas cover a wide range of topics in database systems and information management that include spatial databases, practical machine learning, and data science. Dr. Eunus’s research papers have been published in top-ranking journals and conferences such as the VLDB Journal, TKDE, ICDE, CIKM, PVLDB, and UbiComp. He also served as a Program Committee Member of many prestigious conferences that include SIGMOD, VLDB, AAAI, and SIGSPATIAL. Dr Eunus is a senior member of Association of Computing Machinery (ACM), USA.
\end{IEEEbiography}


%\vfill


\end{document}